\documentclass[aps,preprint,amsmath,amssymb,11pt]{revtex4}
\usepackage{graphicx}
\usepackage{epstopdf}
\pdfoutput=1 
\usepackage[T1]{fontenc}
\newcommand\sss{\scriptscriptstyle}
\newcommand{\mW}{m_{\sss W}}
\newcommand{\sW}{s_{\sss W}}
\newcommand{\cW}{c_{\sss W}}

\begin{document}

\title{Constraints on Higgs effective couplings in $H\nu \bar{\nu}$ production of CLIC at 380 GeV}
\author{H. Denizli}
\email[]{denizli_h@ibu.edu.tr}
\author{A. Senol}
\email[]{senol_a@ibu.edu.tr} 
\affiliation{Department of Physics, Abant Izzet Baysal University, 14280, Bolu, Turkey}
\begin{abstract}
The potential of the $e^+e^-\to\nu \bar{\nu} H$ process in the first stage of CLIC considering center of mass energy of 380 GeV and assuming the baseline integrated luminosity of 500 fb$^{-1}$  is examined to probe CP conserving dimension-six operators in a model-independent Standard Model effective field theory framework. In the analysis, a detailed fast simulation on $e^+e^-\to\nu \bar{\nu} H$ signal processes and dominant backgrounds are performed including parton showering with PYTHIA and detector simulation based on ILD type detector with DELPHES in MadGraph. The obtained best limits on the $\bar{c}_{HB} $, $\bar{c}_{W}=- \bar{c}_{B}$ and $\bar{c}_{HW} $ are $[-4.82;1.53]\times10^{-2}$, $[-5.11;4.13]\times10^{-3}$ and $[-6.58;5.55]\times10^{-3}$, respectively.

\end{abstract}

\maketitle

\section{Introduction}
The recent Large Hadron Collider (LHC) discovery of a scalar particle with 125 GeV  which is compatible with Standard Model (SM) Higgs boson predicted by Brout-Englert-Higgs symmetry breaking mechanism opens up a gateway to search for physics beyond the SM \cite{Aad:2012tfa, Chatrchyan:2012xdj}. But, an evidence for new physics beyond the SM using analysis of combined  ATLAS and CMS data for probing the couplings of Higgs boson has not been observed yet.
 Possible deviation from the SM predictions of Higgs boson couplings would imply the presence of new physics involving massive particles that are decoupled at energy scales much larger than the Higgs sector energies being probed \cite{Appelquist:1974tg}.  The SM Effective Field Theory (EFT) is a well-known model independent method for investigation of any deviation from SM \cite{Buchmuller:1985jz,Grzadkowski:2010es}. The origin of this method is based on all new physics contributions to the SM described by a systematic expansion in a series of high dimensional operators beyond the SM fields. All high dimensional operators conform to $SU(3)_C\times SU(2)_L\times U(1)_Y$ SM gauge symmetry. The dimension-6 operators play an important role in the framework since they match to ultraviolet (UV) models which are simplified by the universal one-loop effective action. There have been many analyses for constraints on SM EFT operators with available data from LHC-Run 1 \cite{Corbett:2012ja,Ellis:2014jta,Ellis:2014dva, Falkowski:2015fla,Corbett:2015ksa,Ferreira:2016jea,Aad:2015tna,Green:2016trm} and with electroweak precision measurements provided from previous accelerator, namely Large Electron Positron (LEP) \cite{Jones:1979bq,Grinstein:1991cd,Hagiwara:1993qt, Han:2004az}. Especially, the prediction on dimension-6 operators have been examined in many rewarding studies at High Luminosity LHC (HL-LHC) \cite{Englert:2015hrx,Buckley:2015lku, Khanpour:2017inb} and future $e^+e^-$ colliders \cite{Amar:2014fpa,Kumar:2015eea,Ellis:2015sca,Ge:2016zro,Cohen:2016bsd,Ellis:2017kfi,Alam:2017hkf,Khanpour:2017cfq,Englert:2017gdy}. 

The precision measurements of Higgs boson couplings with the other SM particles at the LHC and planned future colliders will give us detailed information about its true nature. The future multi-TeV $e^+e^-$ colliders with extremely high luminosity and clean environment due to the absence of hadronic initial state, would give access to precise measurement, especially for the Higgs couplings. The  Compact Linear Collider (CLIC) is one the mature proposed linear colliders with centre of mass energies from a few hundred GeV up to 3 TeV \cite{CLIC:2016zwp}. The first energy stage of CLIC operation was chosen to be $\sqrt s$=380 GeV, with the predicted integrated luminosity of 500 $fb^{-1}. $ The primary motivation of  this stage is the precision measurements of SM Higgs properties and also the model independent Higgs couplings to both fermions and bosons \cite{CLIC:2016zwp, Abramowicz:2016zbo}. 

In this study, we focus on the analysis of $e^+e^-\to \nu \bar{\nu} H$ production process in order to assess the projection of the first energy stage of the CLIC on the CP-conserving dimension-6 operators involving the Higgs and gauge bosons ($W^{\pm}$, $\gamma$, $Z$) defined by an SM EFT Lagrangian in the next section.

\section{Effective Operators}
The well known SM Lagrangian ( $\mathcal{L}_{\rm SM}$ ) involving renormalizable interactions is suppressed by higher dimensional operators in SM EFT approach. All these operators parametrised by an energy scale of non-observed states assumed larger than vacuum expectation value of Higgs field ($v$). A few different operator bases are presented in the literature, we consider SM EFT operators as the strongly interacting light Higgs Lagrangian ($\mathcal{L}_{\rm SILH}$) in bar convention \cite{Englert:2015hrx,Contino:2013kra,Alloul:2013naa}.  Assuming the baryon and lepton number conservation, the most general form of dimension-6 effective Lagrangian including Higgs boson couplings that keep SM gauge symmetry is given as follows;
\begin{eqnarray}
\mathcal{L} = \mathcal{L}_{\rm SM} + \sum_{i}\bar c_{i}O_{i}
\end{eqnarray}
where  $\bar c_{i}$ are normalized Wilson coefficients that are free parameters. In this work, we consider the dimension-6 CP-conserving interactions of the Higgs boson and electroweak gauge boson in SILH basis as \cite{Alloul:2013naa}:
\begin{eqnarray}\label{massb}
	\begin{split}
		\mathcal{L}_{\rm SILH} = & \
		 \frac{\bar c_{H}}{2 v^2} \partial^\mu\big[\Phi^\dag \Phi\big] \partial_\mu \big[ \Phi^\dagger \Phi \big]
		+ \frac{\bar c_{T}}{2 v^2} \big[ \Phi^\dag {\overleftrightarrow{D}}^\mu \Phi \big] \big[ \Phi^\dag {\overleftrightarrow{D}}_\mu \Phi \big] - \frac{\bar c_{6} \lambda}{v^2} \big[\Phi^\dag \Phi \big]^3
		\\
		& \
		  - \bigg[\frac{\bar c_{u}}{v^2} y_u \Phi^\dag \Phi\ \Phi^\dag\cdot{\bar Q}_L u_R
		  + \frac{\bar c_{d}}{v^2} y_d \Phi^\dag \Phi\ \Phi {\bar Q}_L d_R
		+\frac{\bar c_{l}}{v^2} y_l \Phi^\dag \Phi\ \Phi {\bar L}_L e_R
		 + {\rm h.c.} \bigg]
		\\
		&\
		 + \frac{i g\ \bar  c_{W}}{m_{W}^2} \big[ \Phi^\dag T_{2k} \overleftrightarrow{D}^\mu \Phi \big]  D^\nu  W_{\mu \nu}^k + \frac{i g'\ \bar c_{B}}{2 m_{W}^2} \big[\Phi^\dag \overleftrightarrow{D}^\mu \Phi \big] \partial^\nu  B_{\mu \nu} \\
		&\   
		+ \frac{2 i g\ \bar c_{HW}}{m_{W}^2} \big[D^\mu \Phi^\dag T_{2k} D^\nu \Phi\big] W_{\mu \nu}^k  
		+ \frac{i g'\ \bar c_{HB}}{m_{W}^2}  \big[D^\mu \Phi^\dag D^\nu \Phi\big] B_{\mu \nu}   \\
		&\
		 +\frac{g'^2\ \bar c_{\gamma}}{m_{W}^2} \Phi^\dag \Phi B_{\mu\nu} B^{\mu\nu}  
		+\frac{g_s^2\ \bar c_{g}}{m_{W}^2} \Phi^\dag \Phi G_{\mu\nu}^a G_a^{\mu\nu} 
	\end{split}
\end{eqnarray}
where $\lambda$ represents the Higgs quartic coupling; $y_u$, $y_d$ and $y_l$ are the $3\times3$ Yukawa coupling matrices in flavor space; $g'$, $g$ and $g_s$ denotes coupling constant of  $U(1)_Y$, $SU(2)_L$ and $SU(3)_C$ gauge fields, respectively; the generators of $SU(2)_L$ in the fundamental representation are given by $T_{2k}=\sigma_k/2$, $\sigma_k$ being the Pauli matrices; $\Phi$ is Higgs field contains a single $SU(2)_L$ doublet of fields; $B^{\mu\nu}=\partial_\mu B_\nu-\partial_\nu B_\mu$  and $W^{\mu \nu}=\partial_\mu W_\nu^k-\partial_\nu W_\mu^k+g\epsilon_{ijk}W_\mu^iW_\nu^j$ are the electroweak field strength tensors and $G^{\mu\nu}$ is the strong field strength tensors; and the Hermitian derivative operators  defined as, 
$\Phi^\dag \overleftrightarrow{D}_\mu \Phi = \Phi^\dag D^{\mu}\Phi - D_{\mu}\Phi^{\dag}\Phi$.
The SM EFT Lagrangian (Eq.(\ref{massb})) containing the Wilson coefficients in the SILH bases of dimension-6 CP-conserving operators can be defined in terms of the mass eigenstates after electroweak symmetry breaking (Higgs boson, W, Z, photon, etc.) as follows
\begin{eqnarray}\label{gbase}
	\mathcal{L} &=& - \frac{m_{ H}^2}{2 v} g^{(1)}_{ hhh}h^3 + \frac{1}{2} g^{(2)}_{ hhh} h\partial_\mu h \partial^\mu h 
    - \frac{1}{4} g_{ hgg} G^a_{\mu\nu} G_a^{\mu\nu} h
    - \frac{1}{4} g_{ h\gamma\gamma} F_{\mu\nu} F^{\mu\nu} h\nonumber\\
   & -& \frac{1}{4} g_{ hzz}^{(1)} Z_{\mu\nu} Z^{\mu\nu} h
    - g_{ hzz}^{(2)} Z_\nu \partial_\mu Z^{\mu\nu} h
    + \frac{1}{2} g_{ hzz}^{(3)} Z_\mu Z^\mu h
    - \frac{1}{2} g_{ haz}^{(1)} Z_{\mu\nu} F^{\mu\nu} h
    - g_{ haz}^{(2)} Z_\nu \partial_\mu F^{\mu\nu} h\nonumber\\
    &-& \frac{1}{2} g_{ hww}^{(1)} {W^+}^{\mu\nu} {W^-}_{\mu\nu} h
    - \Big[g_{ hww}^{(2)} {W^+}^\nu \partial^\mu {W^-}_{\mu\nu} h + {\rm h.c.} \Big]
    +g (1-\frac12 \bar c_{ H}) \mW {W^-}_\mu  {W+}^\mu h\nonumber\\
& -&\bigg[ 
      \tilde y_u \frac{1}{\sqrt{2}} \big[{\bar u} P_R u\big] h +
      \tilde y_d \frac{1}{\sqrt{2}} \big[{\bar d} P_R d\big] h +
      \tilde y_\ell \frac{1}{\sqrt{2}} \big[{\bar \ell} P_R \ell\big] h
     + {\rm h.c.} \bigg] \ ,
\end{eqnarray}

where $W_{\mu\nu}$, $Z_{\mu\nu}$ and $F_{\mu\nu}$ are the field strength tensors of $W$-boson, $Z$-boson and photon, respectively; $m_H$ represent the mass of the Higgs boson;  the effective couplings in gauge basis defined as dimension-6 operators are given in Table~\ref{mtable} in which $a_{H}$ ($g_H$) coupling is the SM contribution to the Higgs boson to two photons (gluons) vertex at loop level.
\begin{table}[h]
\caption{The relations between Lagrangian  parameters in the mass basis (Eq.\ref{massb}) and the Lagrangian in gauge  basis (Eq.\ref{gbase}). ($c_W\equiv\cos \theta_W$, $s_W\equiv\sin \theta_W$)}  
\begin{ruledtabular}\label{mtable}
\begin{tabular}{ll}
    $g_{hhh}^{(1)}$= $1 + \frac78 \bar c_{ 6} - \frac12 \bar c_{H}$&$g_{hhh}^{(2)}$= $\frac{g}{\mW} \bar c_{H}$\\
    $g_{hgg}$= $g_{ H}- \frac{4 \bar c_{g} g_s^2 v}{\mW^2}$ &$g_{h\gamma\gamma}$= $a_{ H} - \frac{8 g \bar c_{ \gamma} \sW^2}{\mW}$ \\
    $g^{(1)}_{ hzz}$= $\frac{2 g}{\cW^2 \mW} \Big[ \bar c_{HB} \sW^2 - 4 \bar c_{ \gamma} \sW^4 + \cW^2 \bar c_{ HW}\Big]$& $g^{(2)}_{ hzz}$= $\frac{g}{\cW^2 \mW} \Big[(\bar c_{ HW} +\bar c_{ W}) \cW^2  + (\bar c_{ B} + \bar c_{ HB}) \sW^2 \Big]$ \\
    $g^{(3)}_{hzz}$=  $\frac{g \mW}{\cW^2} \Big[ 1 -\frac12 \bar c_{H} - 2 \bar c_{T} +8 \bar c_{\gamma} \frac{\sW^4}{\cW^2}  \Big]$&  $g^{(1)}_{ haz}$= $\frac{g \sW}{\cW \mW} \Big[  \bar c_{ HW} - \bar c_{HB} + 8 \bar c_{ \gamma} \sW^2\Big]$ \\
    $g^{(2)}_{haz}$= $\frac{g \sW}{\cW \mW} \Big[  \bar c_{ HW} - \bar c_{ HB} - \bar c_{ B} + \bar c_{ W}\Big]$& $\tilde y_d$= $y_d \Big[1  -\frac12 \bar c_{ H} + \frac32 \bar c_{ d}\Big]$   \\
       $g^{(1)}_{ hww}$= $\frac{2 g}{\mW} \bar c_{ HW}$ &   $\tilde y_u$= $y_u \Big[1 -\frac12 \bar c_{ H} + \frac32 \bar c_{ u}\Big]$ \\
$g^{(2)}_{ hww}$= $\frac{g}{\mW} \Big[ \bar c_{ W} + \bar c_{ HW} \Big]$ & $\tilde y_\ell$= $y_\ell \Big[1  -\frac12 \bar c_{ H} + \frac32 \bar c_{ \ell}\Big]$ \\ 
  \end{tabular}
\end{ruledtabular}
\end{table}
 \begin{figure}
\includegraphics{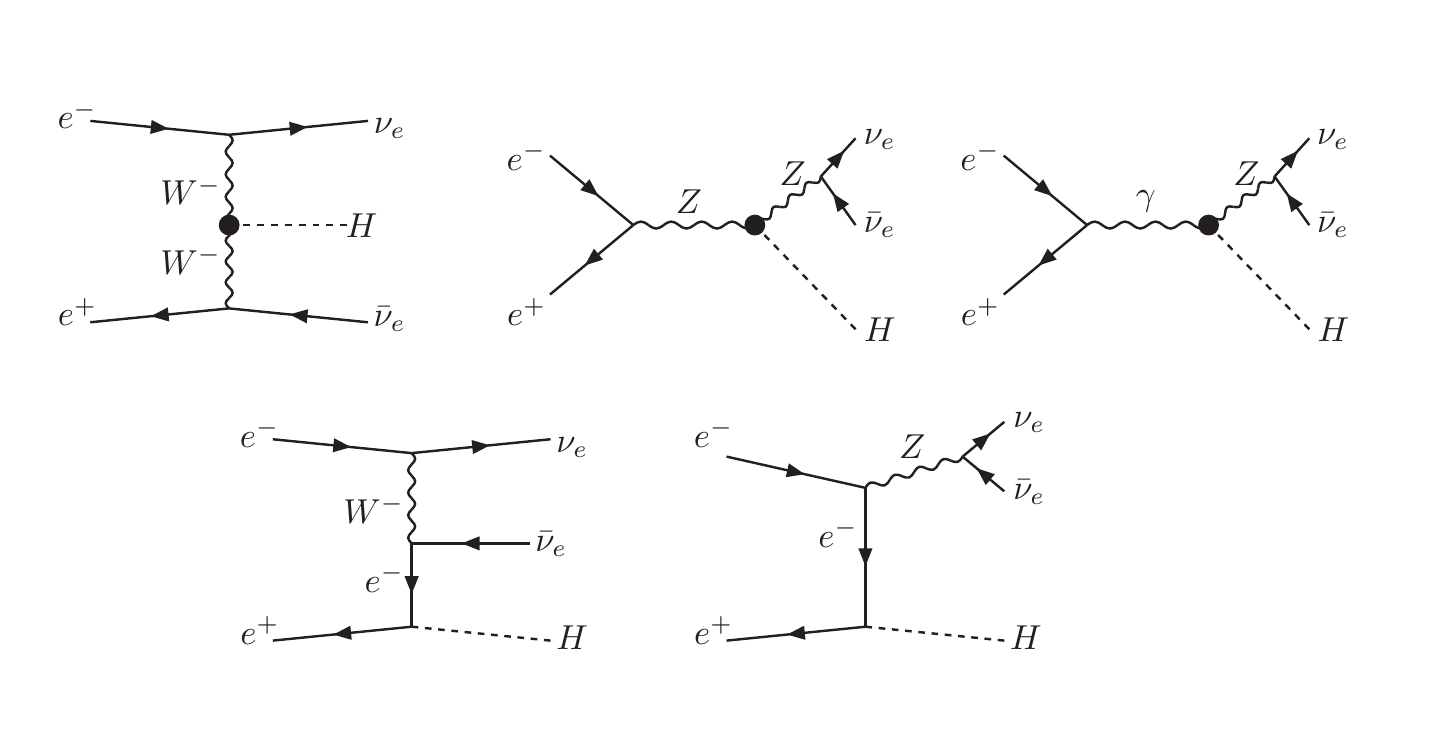} 
\caption{The Feynman diagrams for the process $e^+e^-\to \nu \bar{\nu} H$. \label{fd}}
\end{figure}
\begin{figure}
\includegraphics{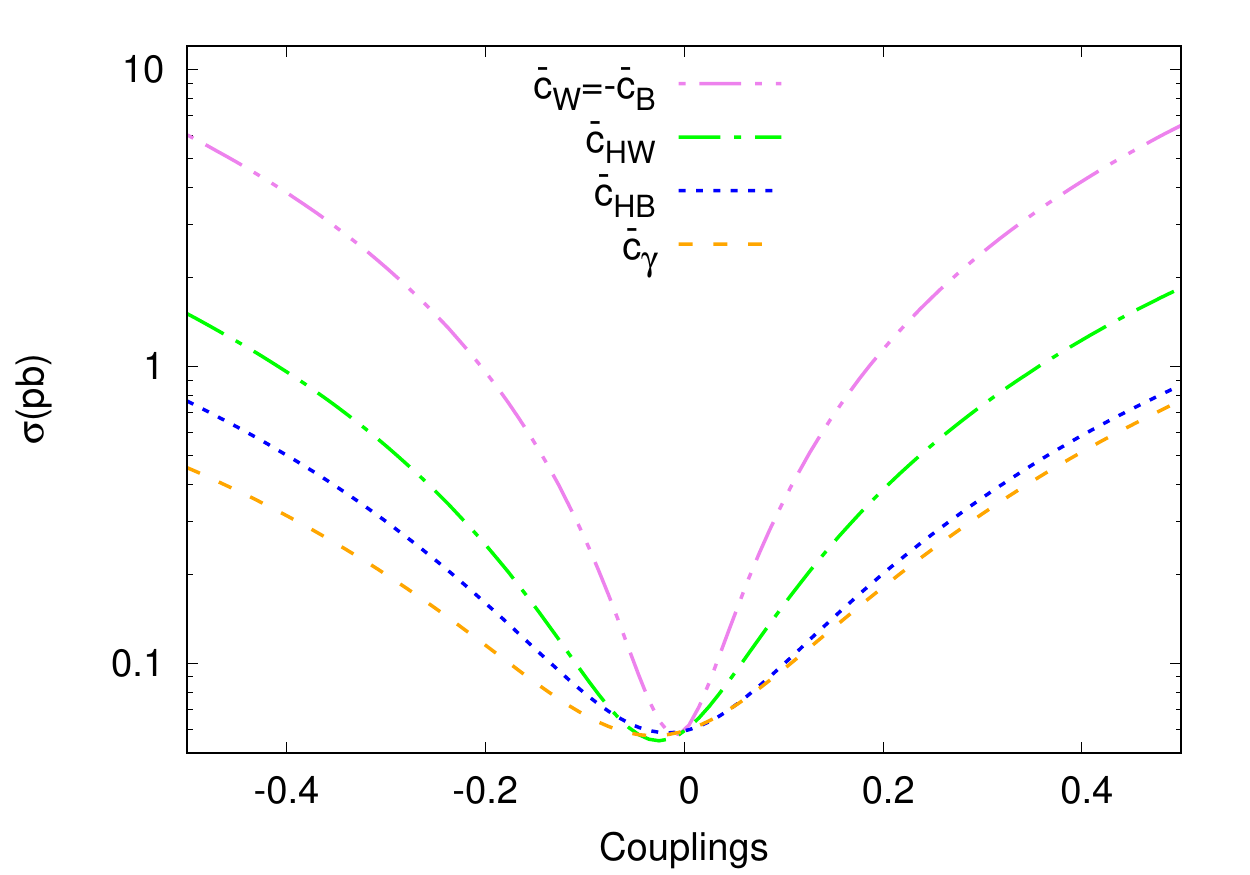} 
\caption{ The total cross section as a function of CP-conserving $\bar{c}_W$=-$\bar{c}_B$, $\bar{c}_{HW}$, $\bar{c}_{HB}$ and $\bar{c}_{\gamma}$ couplings for $e^+e^-\to \nu \bar{\nu} H$ process at the CLIC with $\sqrt s$=380 GeV. \label{fig1}}
\end{figure} 

We use the parametrization in Ref. \cite{Alloul:2013naa}  based on the formulation given in Ref. \cite{Contino:2013kra} in our analysis. The parametrization is not complete as described in detail in section 3 of Ref. \cite{Alonso:2013hga} and also Ref.\cite{Brivio:2017bnu}. It chooses to remove two fermionic invariants while retaining all the bosonic operators. This choice assumes completely unbroken $U(3)$ flavor symmetry of the UV theory where the coefficient of these operators are unit matrices in flavor space. Therefore, we assume flavor diagonal dimension-six  effects. It is sufficient for the purpose of this paper in which we do not consider higher order electroweak effects but only claim a sensitivity study for $\bar c_{ W}$,  $\bar c_{ B}$, $\bar c_{ HW}$, $\bar c_{ HB}$ and $\bar c_{\gamma}$ couplings.

We have used the Monte Carlo simulations with leading order in \verb|MadGraph5_aMC@NLO| \cite{Alwall:2014hca} involving effect of the dimension-6 operators on $H \nu \bar{\nu} $ production mechanism in $e^+e^-$ collisions. The effective Lagrangian of the SM EFT in Eq.(\ref{massb})  is implemented into the \verb|MadGraph5_aMC@NLO| based on FeynRules \cite{Alloul:2013bka} and UFO \cite{Degrande:2011ua} framework.
 In this study, we focus on searching for the dimension-6 Higgs-gauge boson couplings via $e^+e^-\to\nu \bar{\nu} H$ process as shown in Fig.\ref{fd}. 
This process is sensitive to Higgs-gauge boson couplings; $g_{hzz}$, $g_{hww}$, $g_{hz\gamma}$,  and the couplings of a quark or lepton pair and one single Higgs field; $\tilde y_u$, $\tilde y_d$, $\tilde y_l$ in the mass basis.  In the gauge basis, $e^+e^-\to\nu \bar{\nu} H$ process is sensitive to the seven Wilson coefficients:  $\bar c_{ W}$,  $\bar c_{ B}$, $\bar c_{ HW}$, $\bar c_{ HB}$,  $\bar c_{H}$, $\bar c_{ \gamma}$ and  $\bar c_{T}$ related to Higgs-gauge boson couplings and also effective fermionic couplings. Due to the small Yukawa couplings of the first and second generation fermions, we neglect the effective fermionic couplings. We set $\bar c_{ W} + \bar c_{ B}$  and  $\bar c_{T}$ to zero in all our calculations since the linear combination of  $\bar c_{ W} + \bar c_{ B}$  and  $\bar c_{T}$ strongly constrained from the electroweak precision test of the oblique parameters $S$ and $T$.  The cross sections of $e^+e^-\to\nu \bar{\nu} H$ process as a function of $\bar c_{ W}$,  $\bar c_{ B}$, $\bar c_{ HW}$, $\bar c_{ HB}$ and $\bar c_{\gamma}$ couplings are shown in Fig.\ref{fig1}. There have been many studies in the literature considering individual, subsets or simultaneous change of dimension-6 operators \cite{Ellis:2015sca, Ellis:2017kfi}. Here we vary individually dimension- 6 operators and calculate the contributions to the corrections from new physics in the analysis. We presume that only one of the effective couplings is non-zero at any given time, while the other couplings are fixed to zero. One can easily see the deviation from SM for this couplings even in a small value region for $e^+e^-\to\nu \bar{\nu} H$ process. Therefore, we will only consider these five among the Higgs-gauge boson effective couplings in the detailed analysis including detector simulations through the process $e^+e^-\to\nu \bar{\nu} H$ at CLIC with 380 GeV center of mass energy in the next section.

 \section{Signal and Background Analysis}
 We perform the detailed analysis of $\bar c_{ W}$,  $\bar c_{ B}$, $\bar c_{ HW}$, $\bar c_{ HB}$ and $\bar c_{\gamma}$ effective couplings via $e^+e^-\to\nu \bar{\nu} H$ process as well as other relevant background at the first energy stage of CLIC. The  $e^+e^-\to\nu \bar{\nu} H$ signal process includes both s-channel $e^+e^-\to Z^*/\gamma^* \to ZH \to \nu \bar{\nu} H$ process (Higgsstrahlung) and t-channel $e^+e^-\to \nu\bar{\nu} W^*/W^*\to \nu \bar{\nu} H$ process (WW-fusion) as shown in Fig.\ref{fd}. In the initial energy stage of CLIC at $\sqrt s$=380 GeV, these two process have approximately the same amount contribution to the production cross section of the $e^+e^-\to\nu \bar{\nu} H$ process. In our analysis, we include effective dimension-6 couplings and SM contribution as well as interference between effective couplings and SM contributions ($S+B_H$) that lead to $e^+e^-\to\nu \bar{\nu} H$ process where Higgs decay to pair of $b$-quark. We consider the following relevant backgrounds; $B_H$: $e^+e^-\to\nu \bar{\nu} H$ process which has the same final state of the considered signal process including only SM contribution where the Higgs decays to pair of $b$-quark; $B_{ZZ}$: $e^+e^-\to Z Z$ process where one $Z$ decays to $b\bar b$ and the other decays to $\nu\bar \nu$; $B_{tt}$: $e^+e^-\to t \bar t$ process where two b quarks are from $t (\bar t)$ decaying to $W^+b (W^{-} \bar b)$ in which $W^{\pm}$ decay leptonically;  $B_{Z\nu\nu}$: $e^+e^-\to\nu \bar{\nu} Z$ process in which  $Z$ decays to $b\bar b$. The generated signal and all backgrounds at parton level in \verb|MadGraph5_aMC@NLO| are passed through the Pythia 6 \cite{Sjostrand:2006za} for parton shower and hadronization. The detector responses are taken into account with ILD detector card \cite{Behnke:2013lya} in \verb|Delphes 3.3.3| \cite{deFavereau:2013fsa} package. Then, all events are analysed by using the ExRootAnalysis utility \cite{exroot} with ROOT \cite{Brun:1997pa}. 
 
\begin{figure}
\includegraphics[scale=0.16]{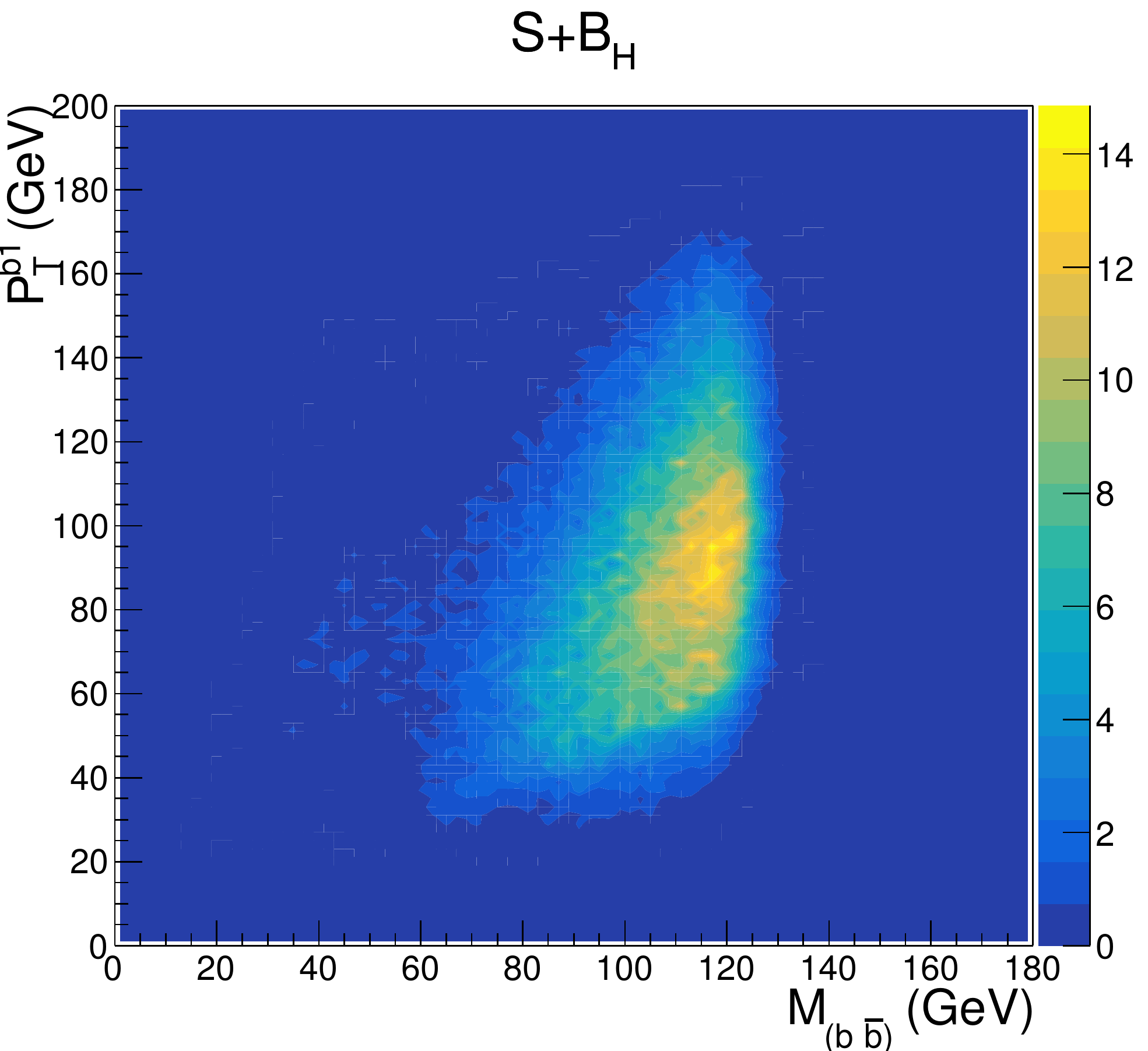}
\includegraphics[scale=0.16]{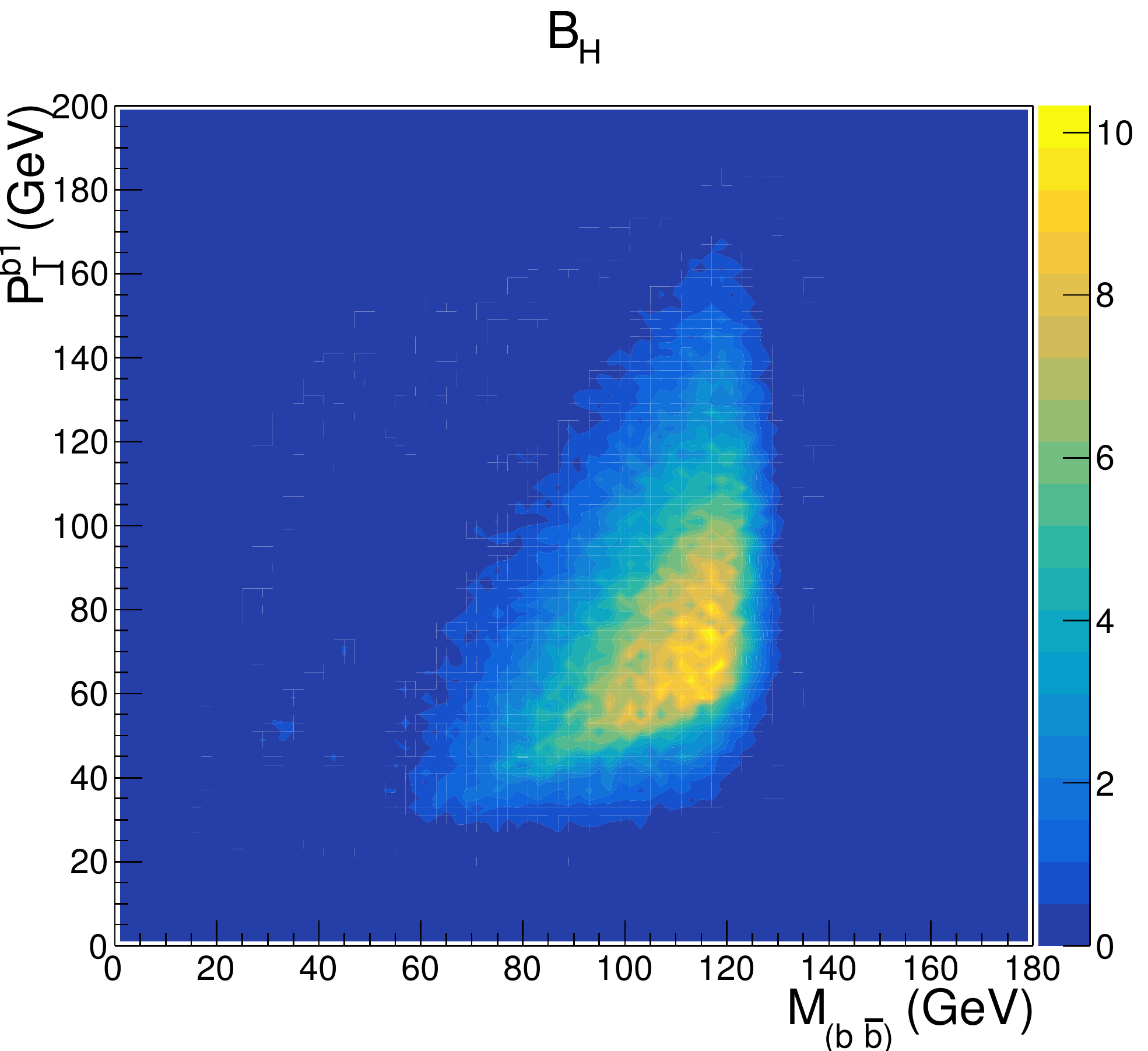}
\includegraphics[scale=0.16]{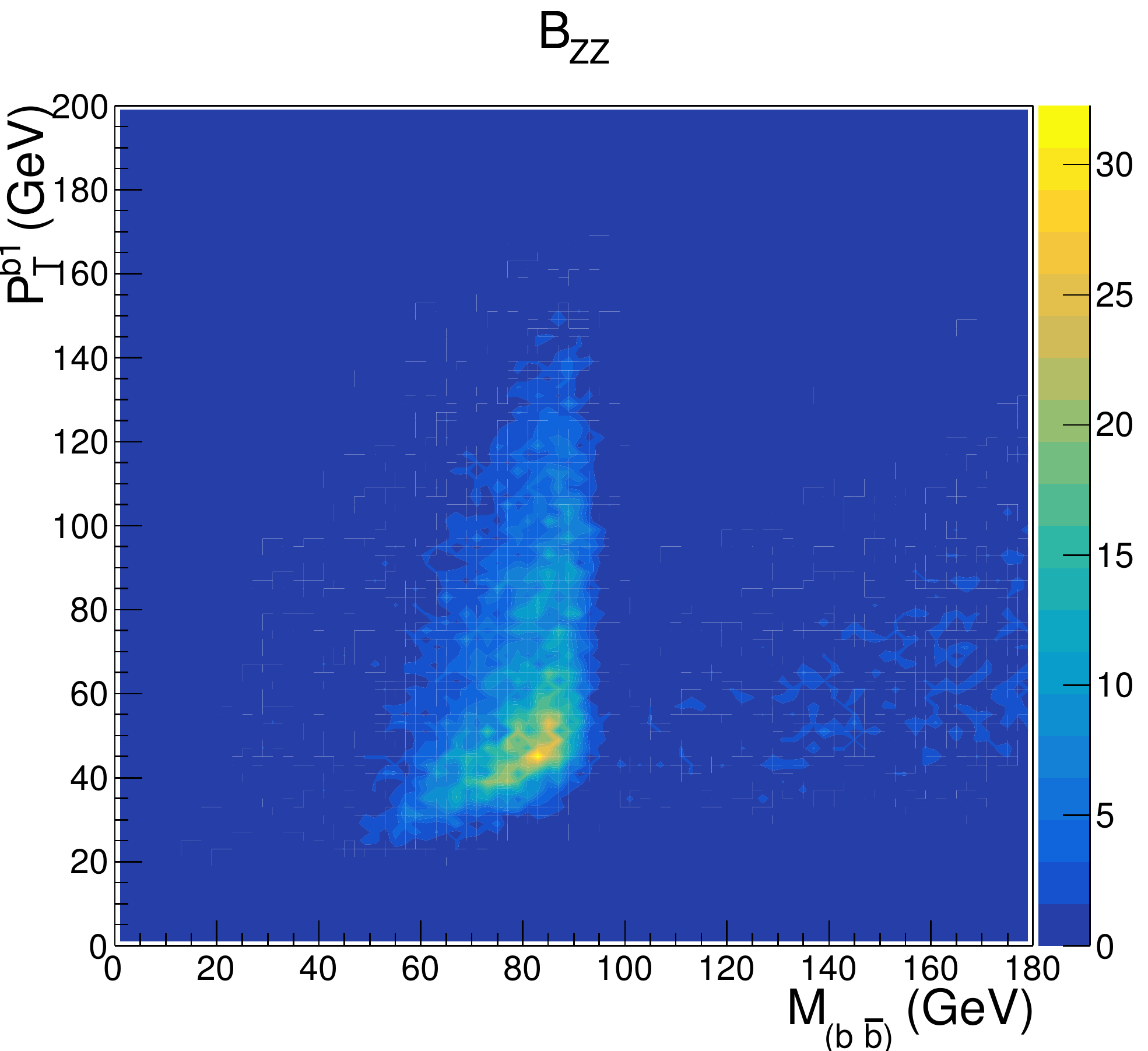}
\includegraphics[scale=0.16]{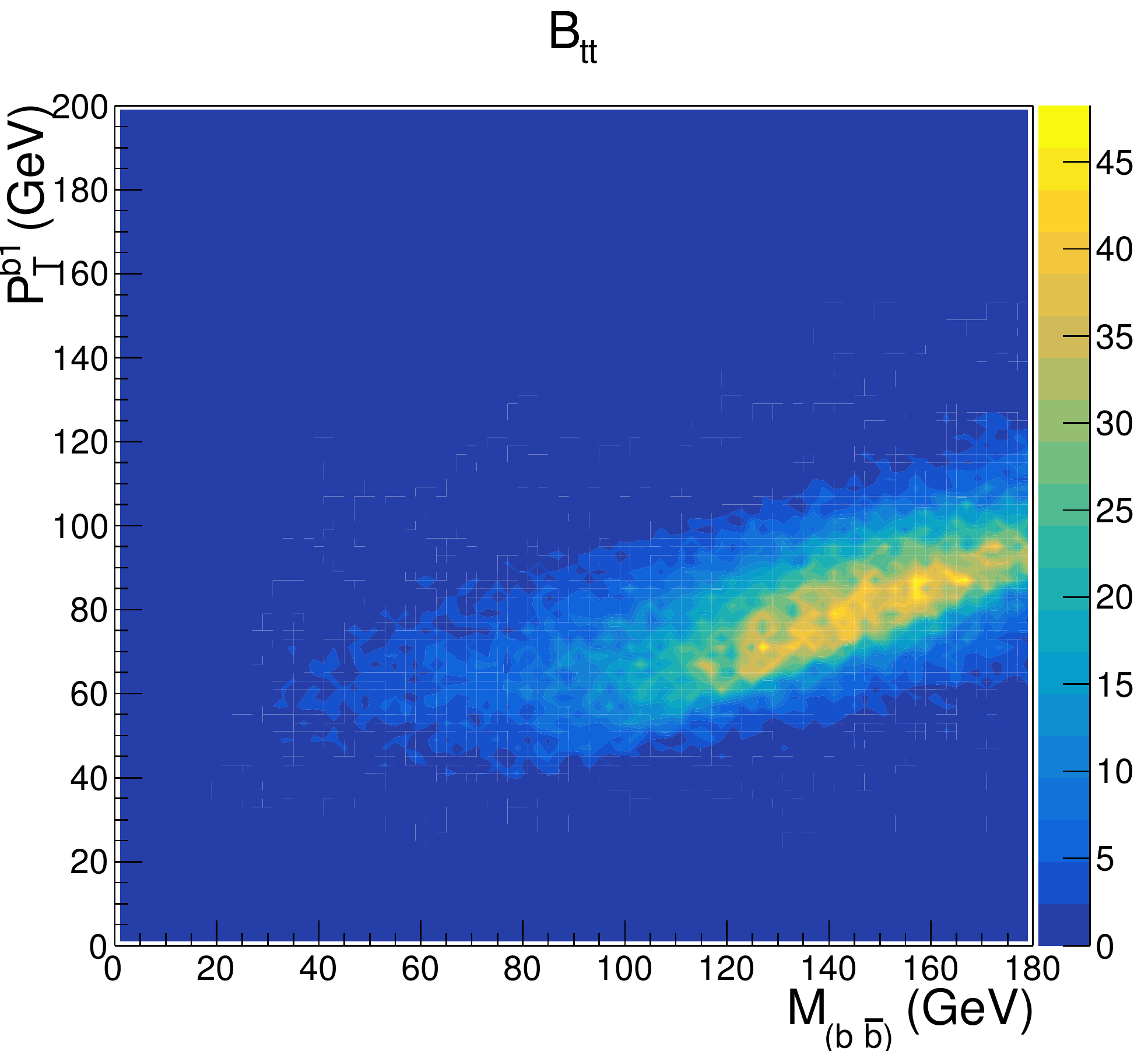}
\includegraphics[scale=0.16]{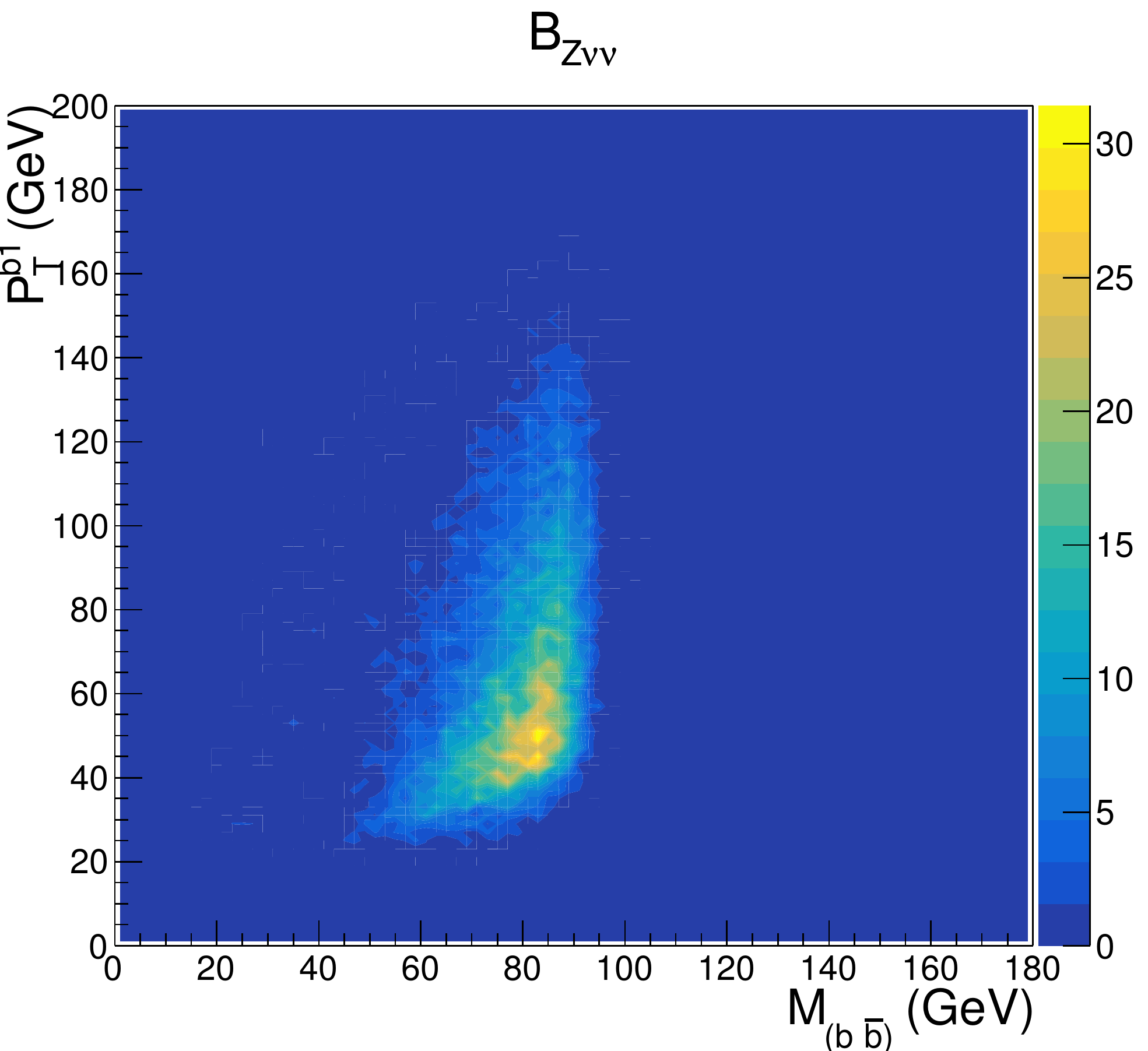}\\
\includegraphics[scale=0.16]{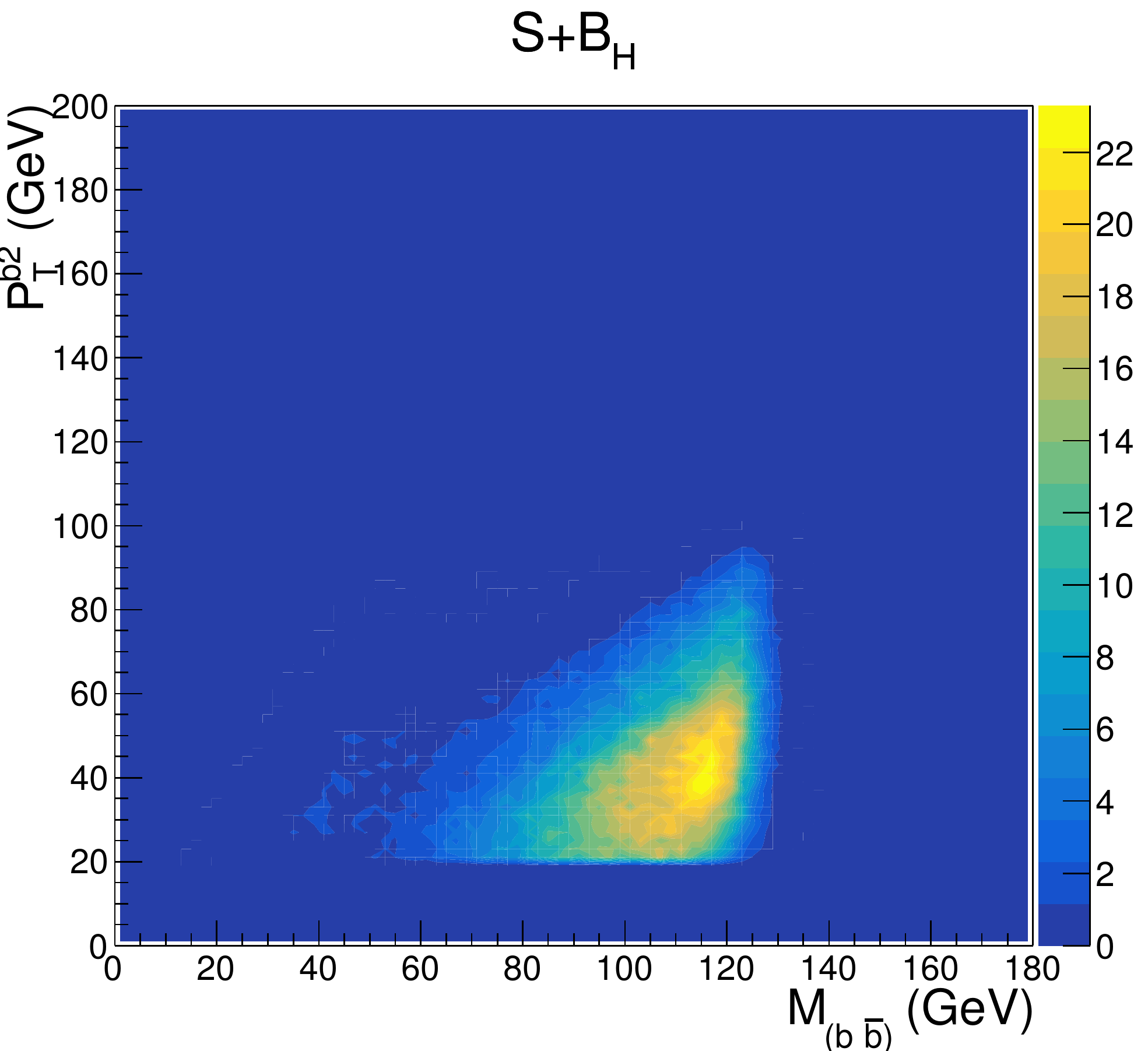}
\includegraphics[scale=0.16]{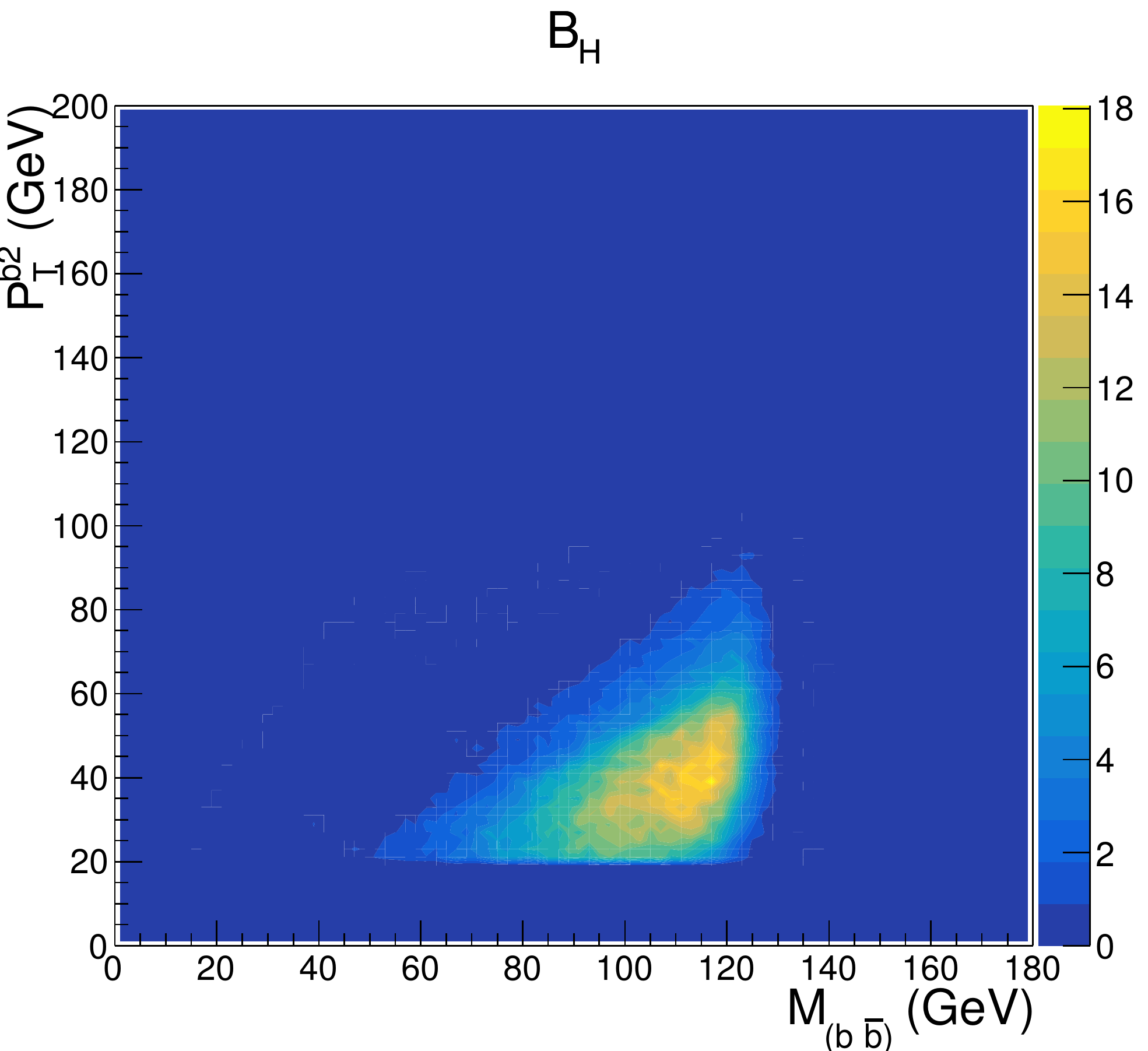}
\includegraphics[scale=0.16]{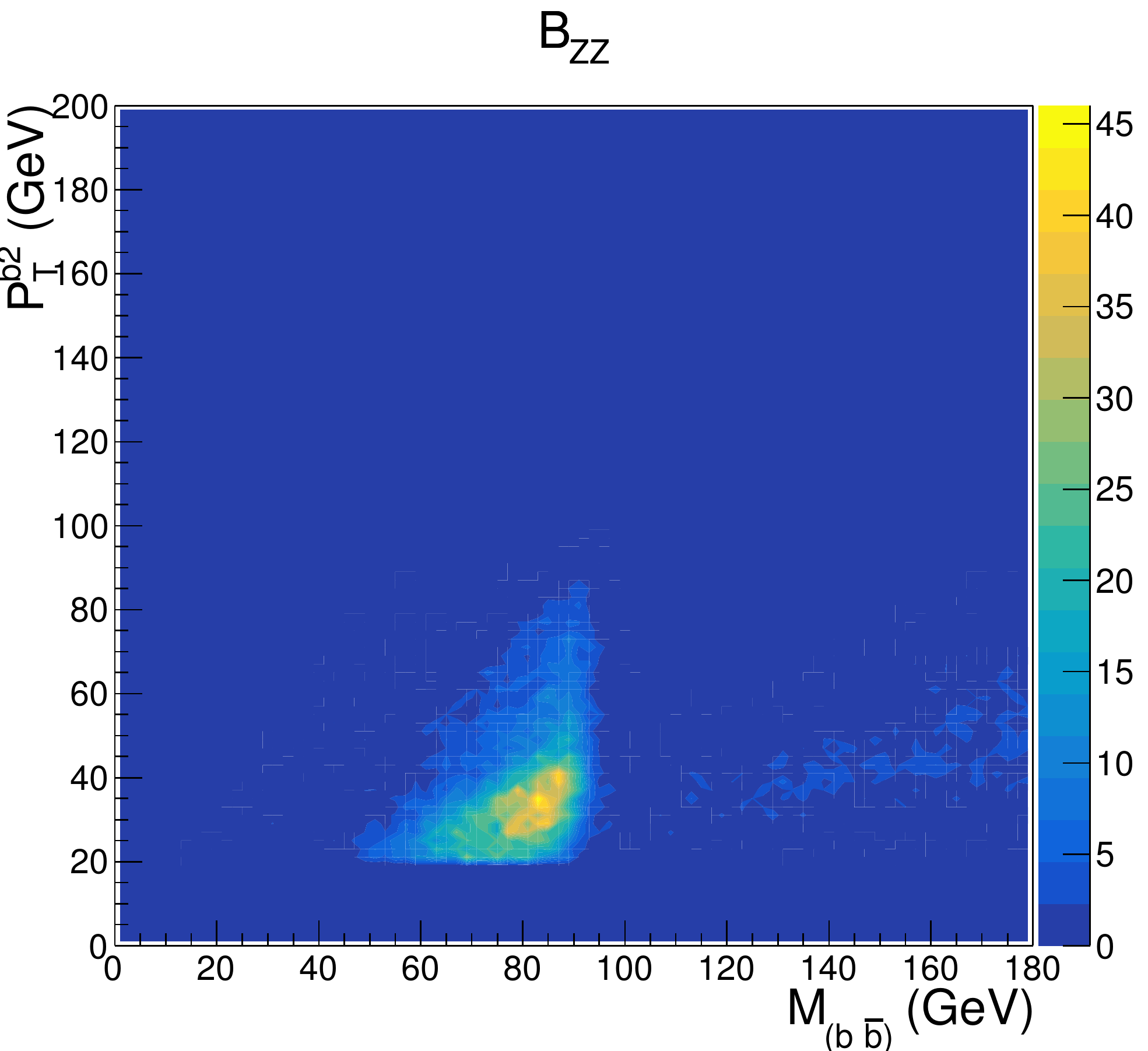}
\includegraphics[scale=0.16]{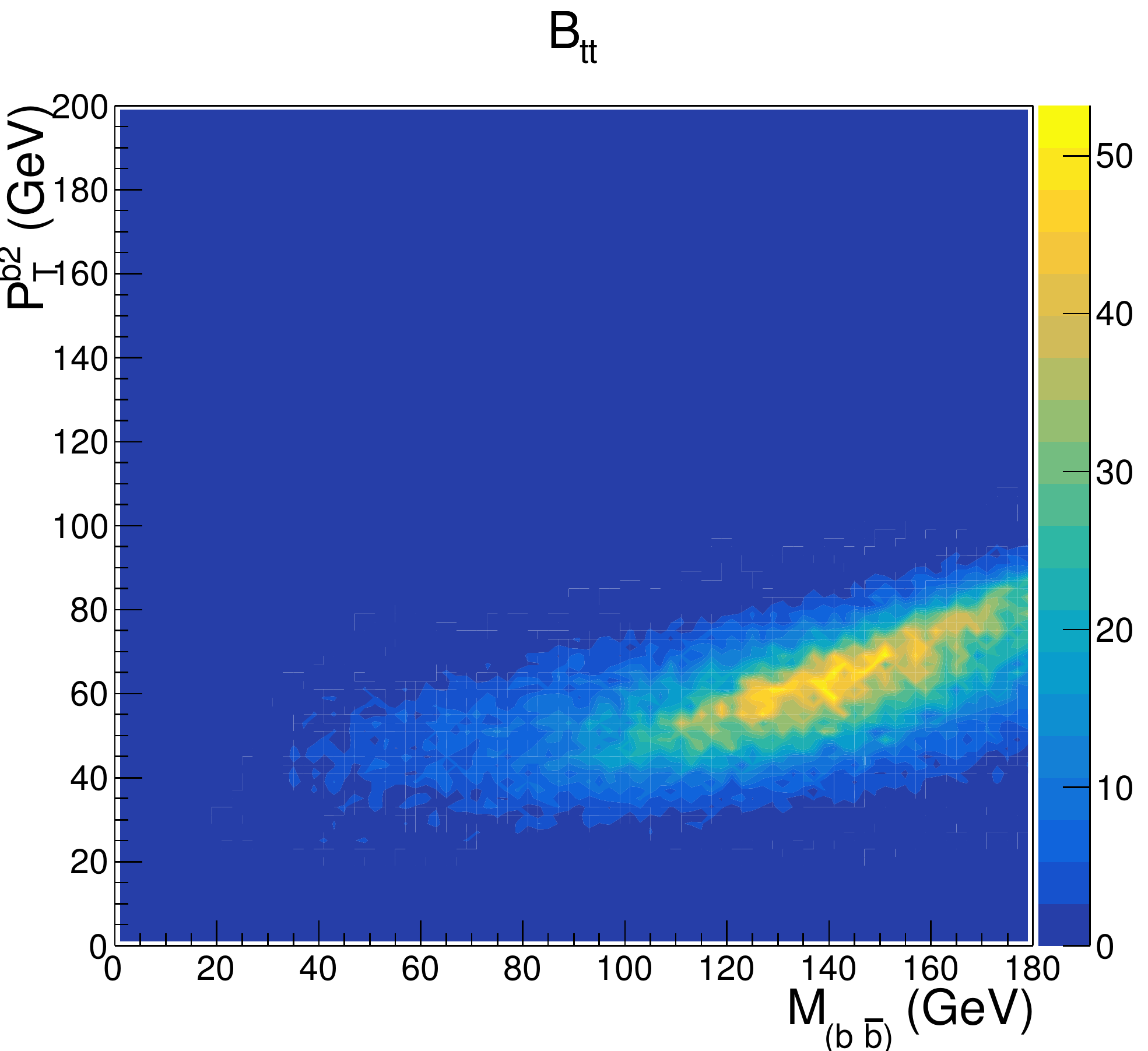}
\includegraphics[scale=0.16]{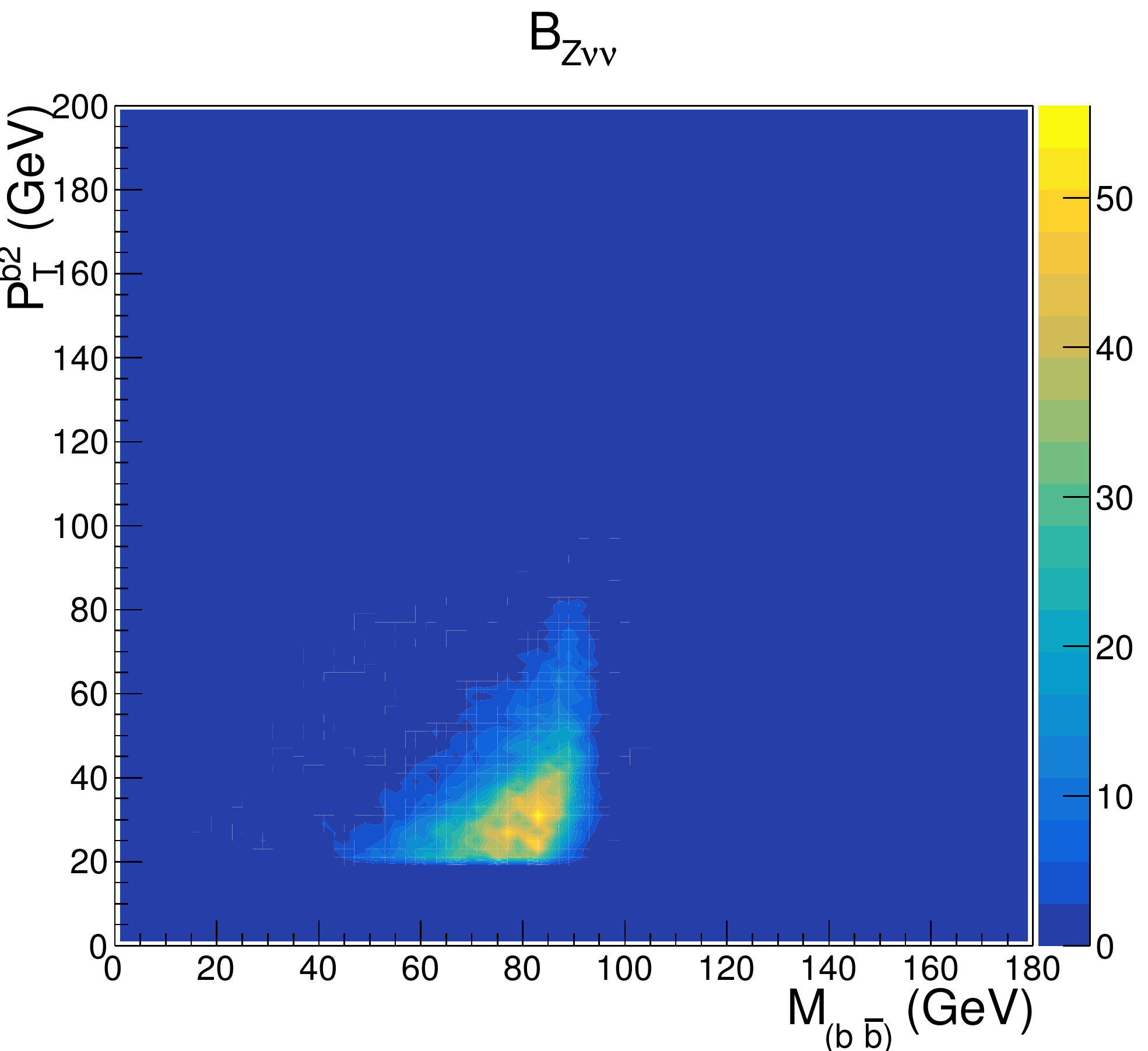}
\caption{ Normalized distributions of transverse momentum of tagged b-jets; $b1$  (first row) and $b2$ (second row) versus reconstructed Higgs boson-mass from $b1$ and $b2$ ($M_{b,\bar b}$) for signal with $\bar{c}_{HW} $=0.05 and relevant background processes.  \label{fig2}}
\end{figure}

\begin{figure}
\includegraphics[scale=0.16]{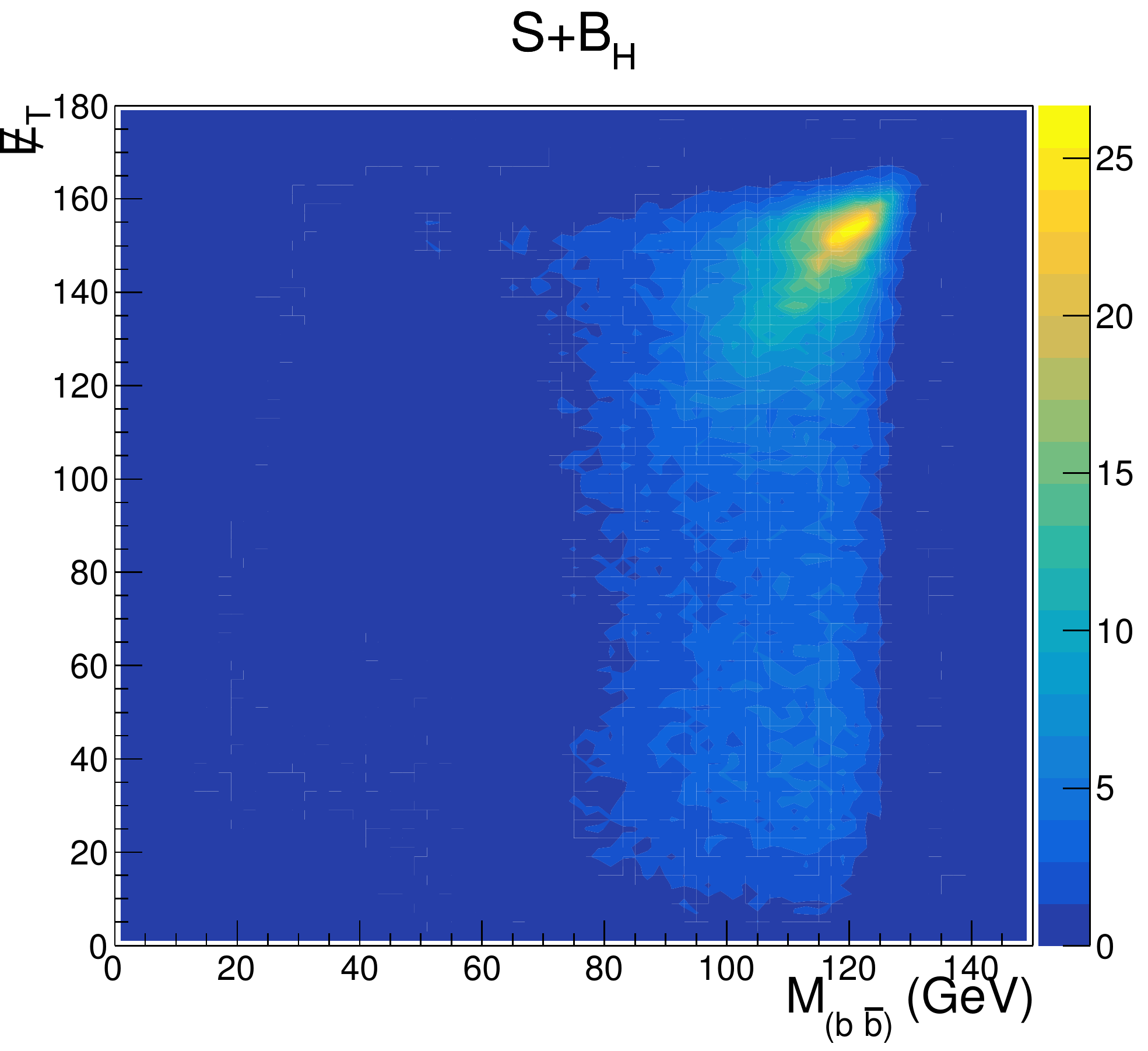}
\includegraphics[scale=0.16]{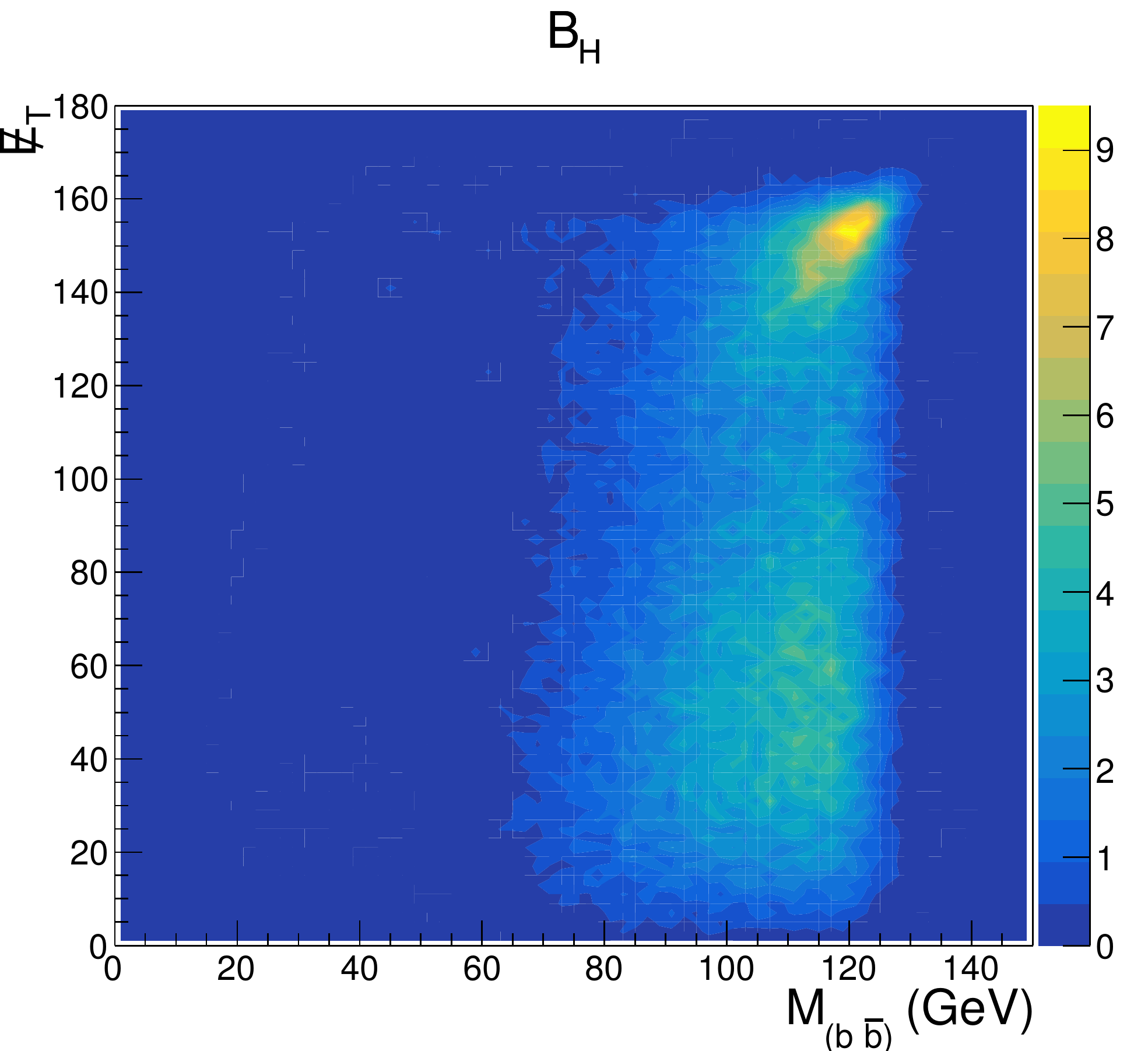}
\includegraphics[scale=0.16]{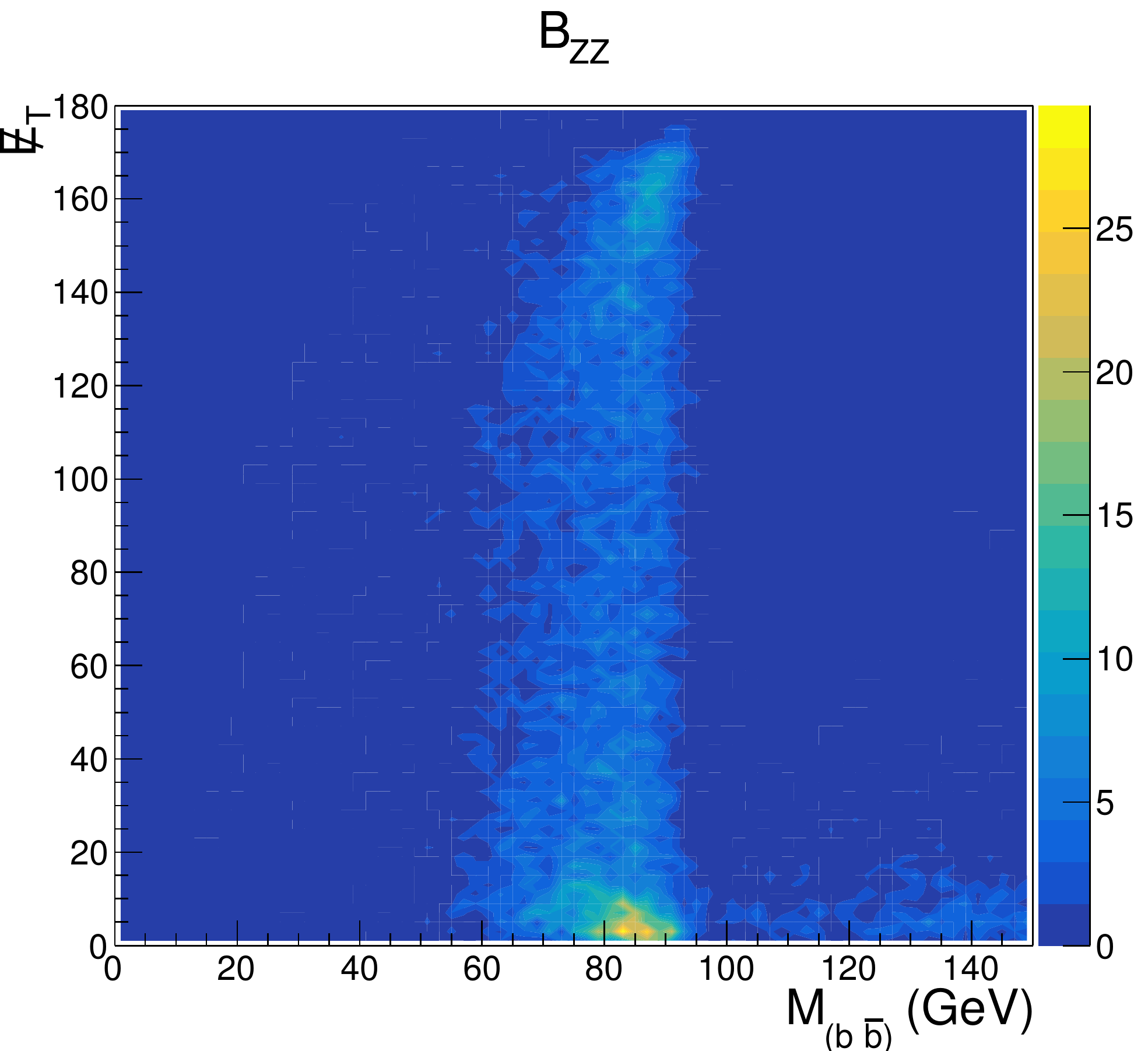}
\includegraphics[scale=0.16]{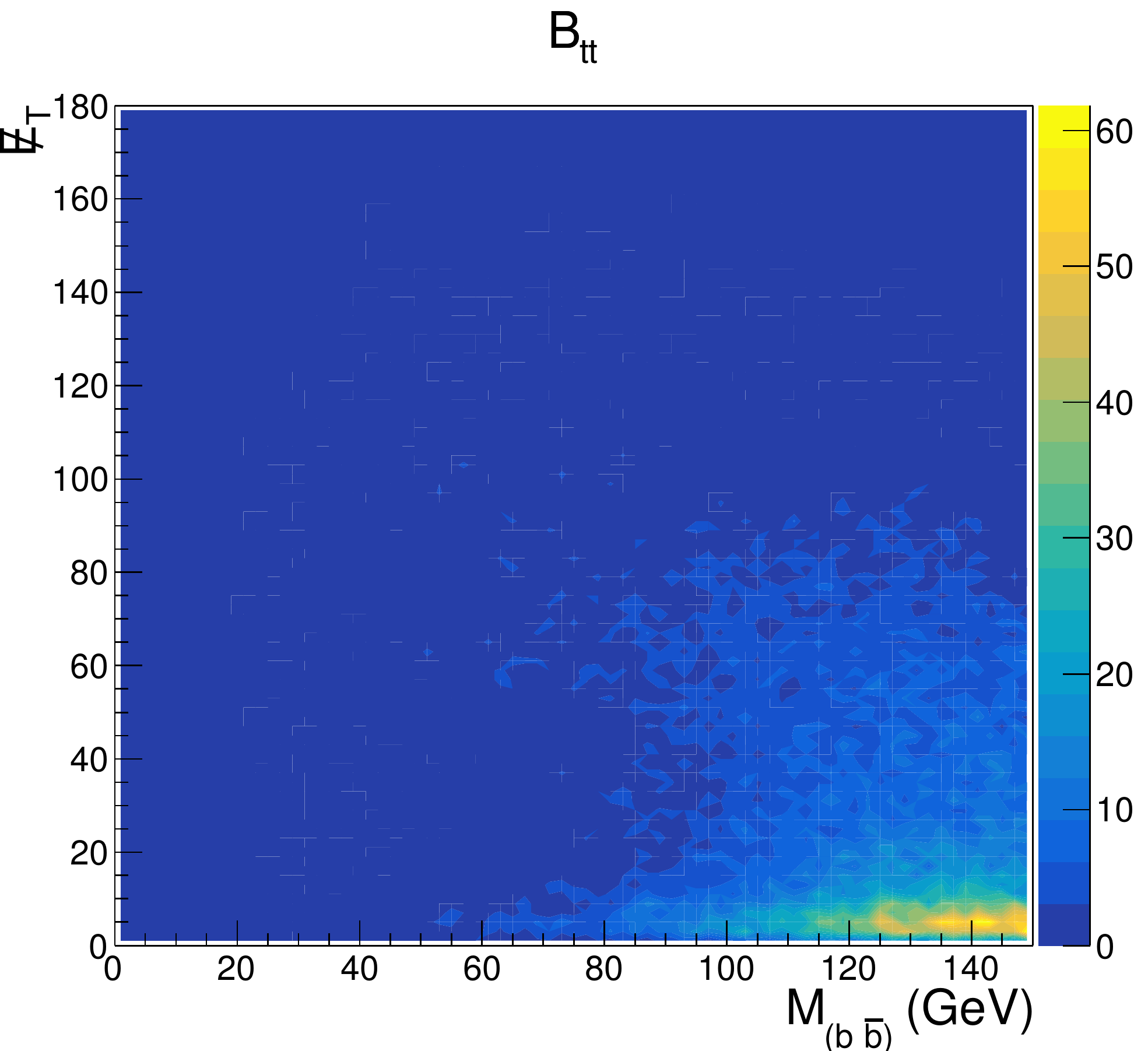}
\includegraphics[scale=0.16]{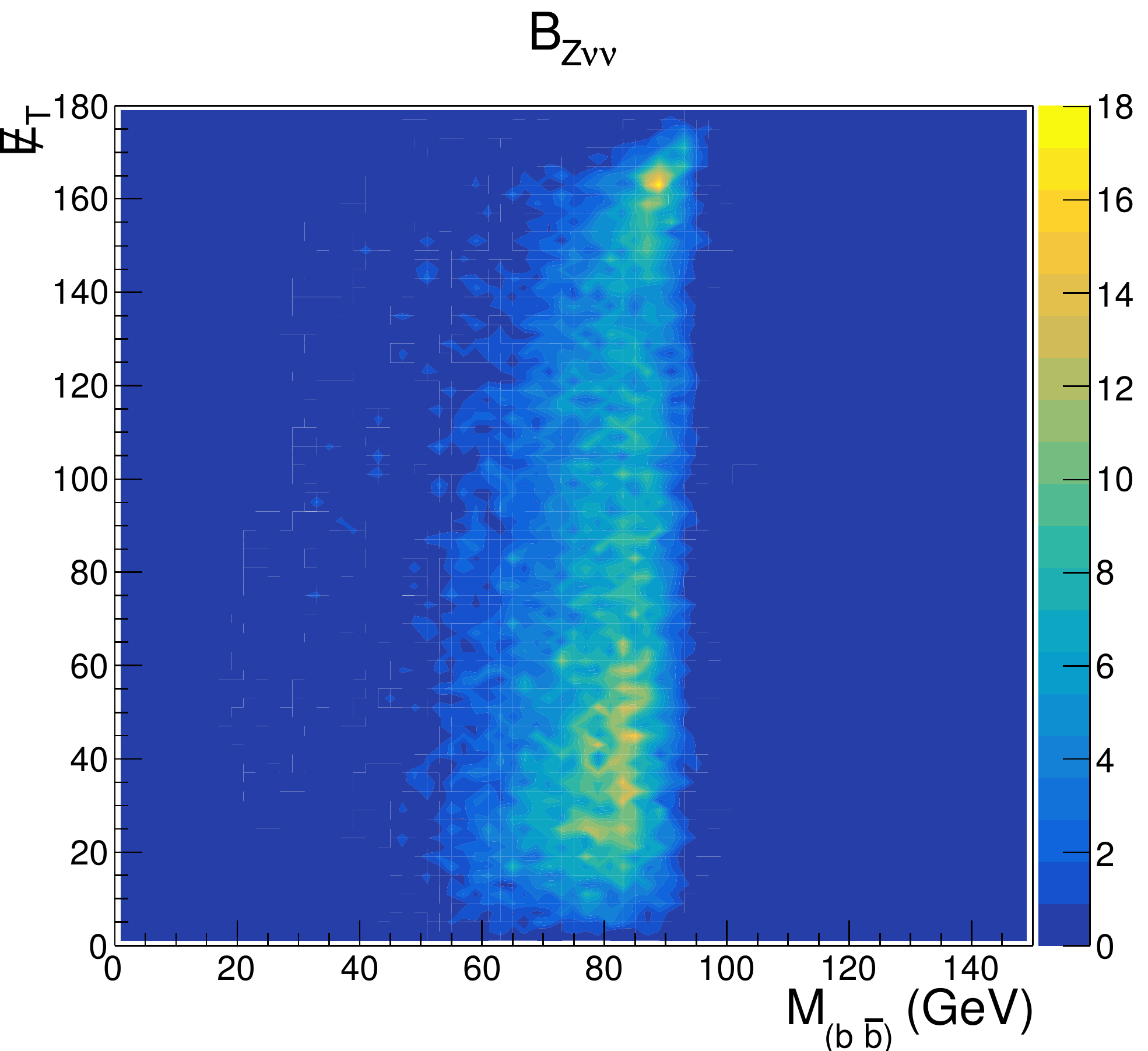}\\
\includegraphics[scale=0.16]{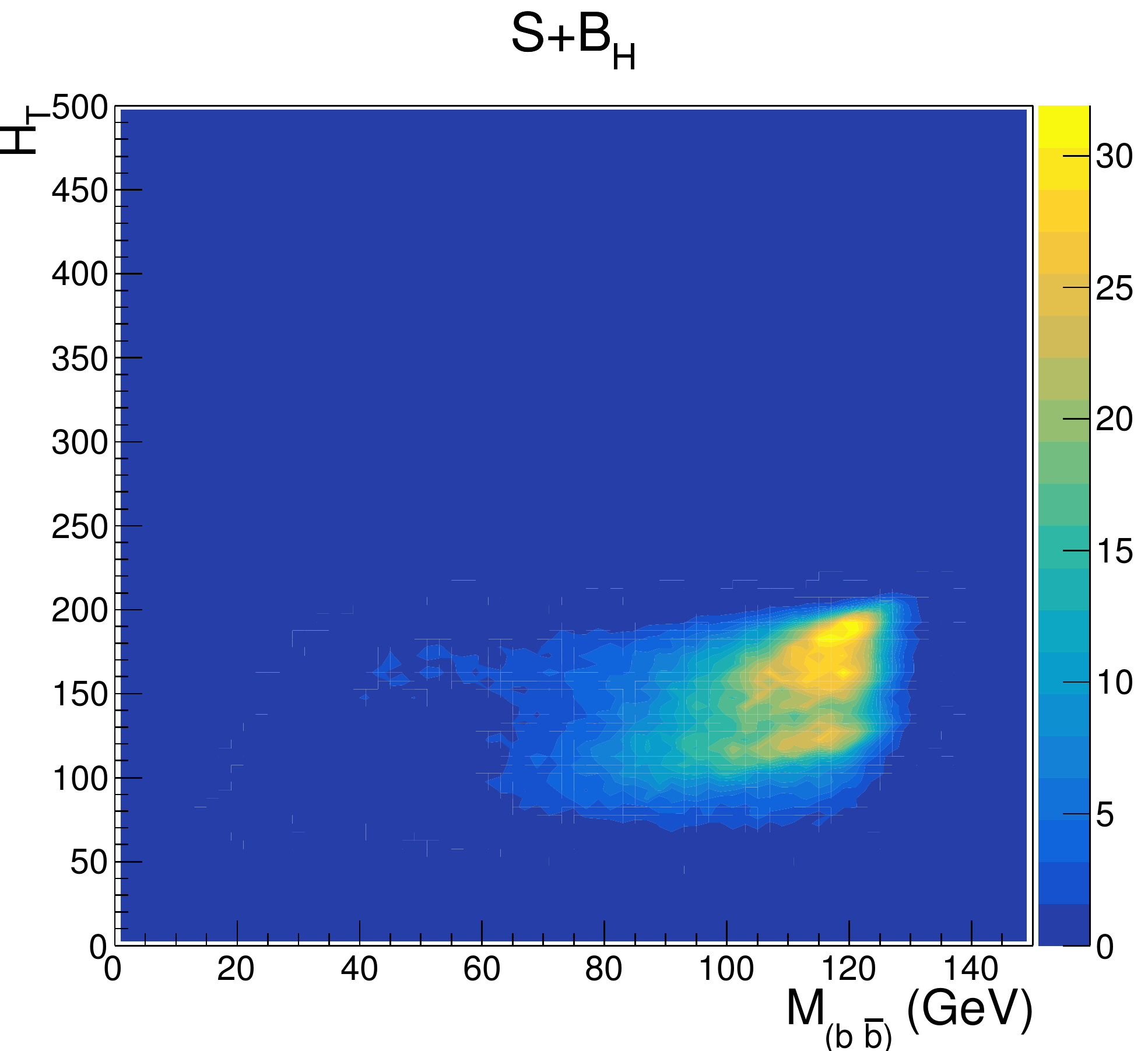}
\includegraphics[scale=0.16]{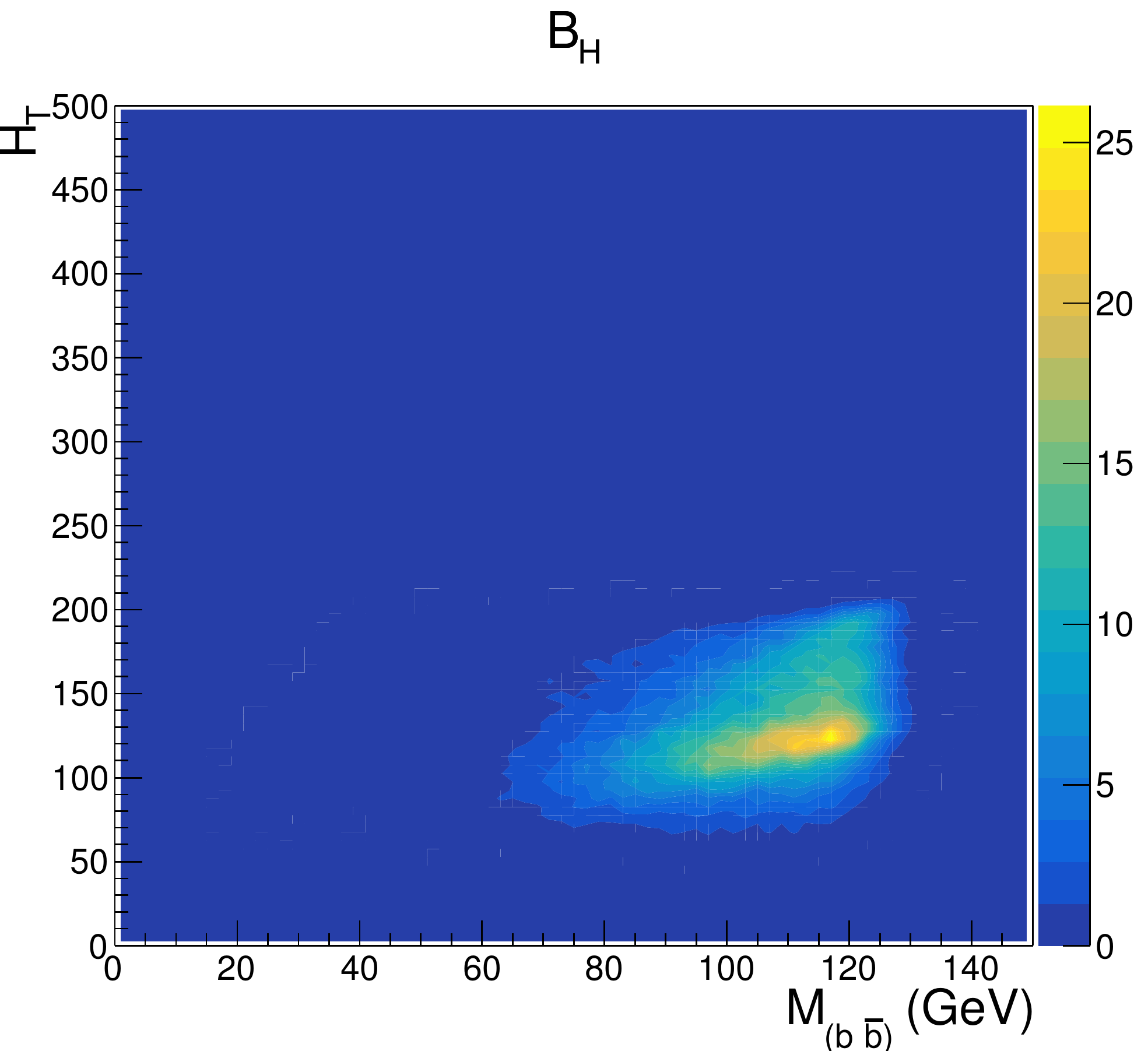}
\includegraphics[scale=0.16]{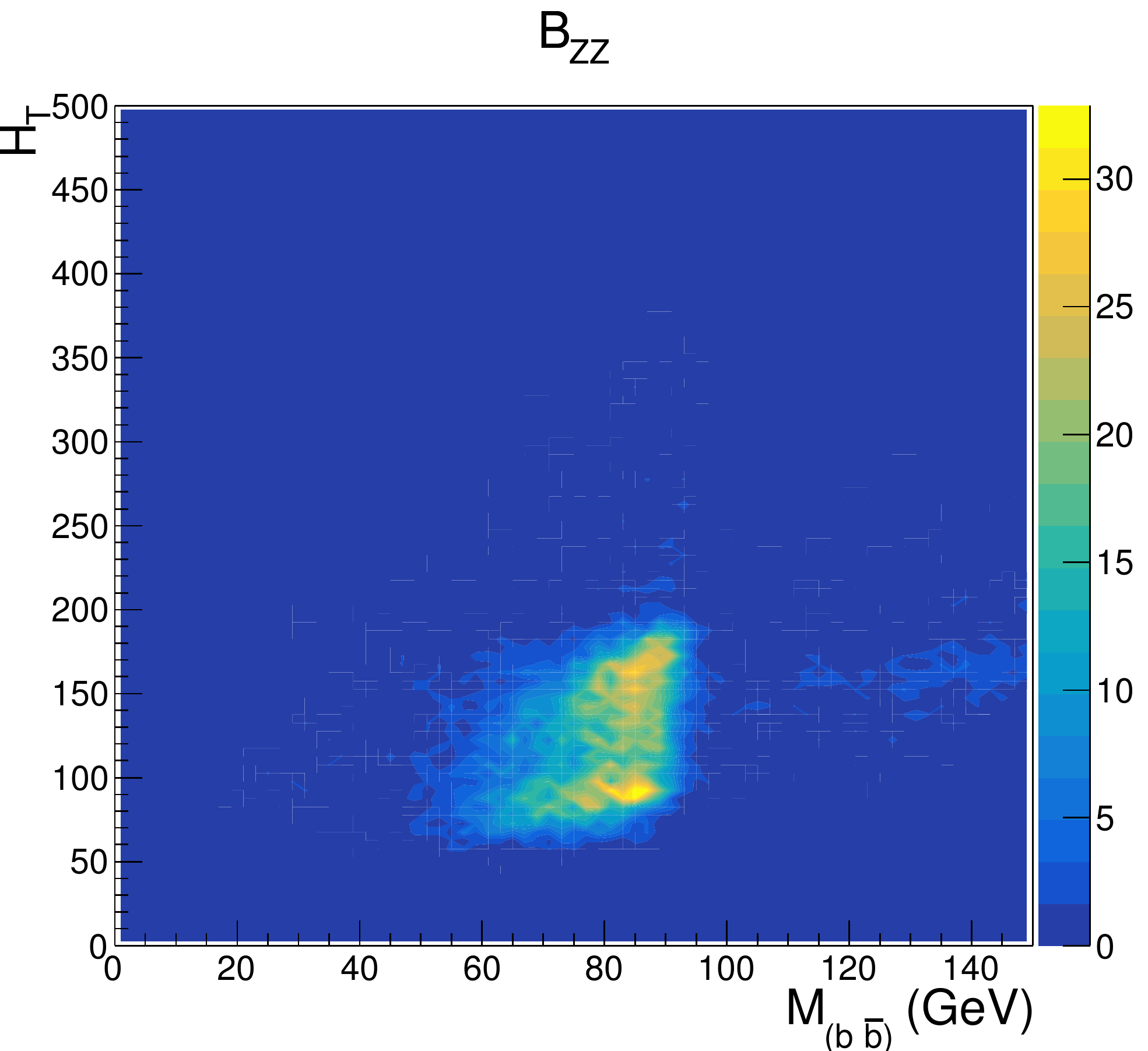}
\includegraphics[scale=0.16]{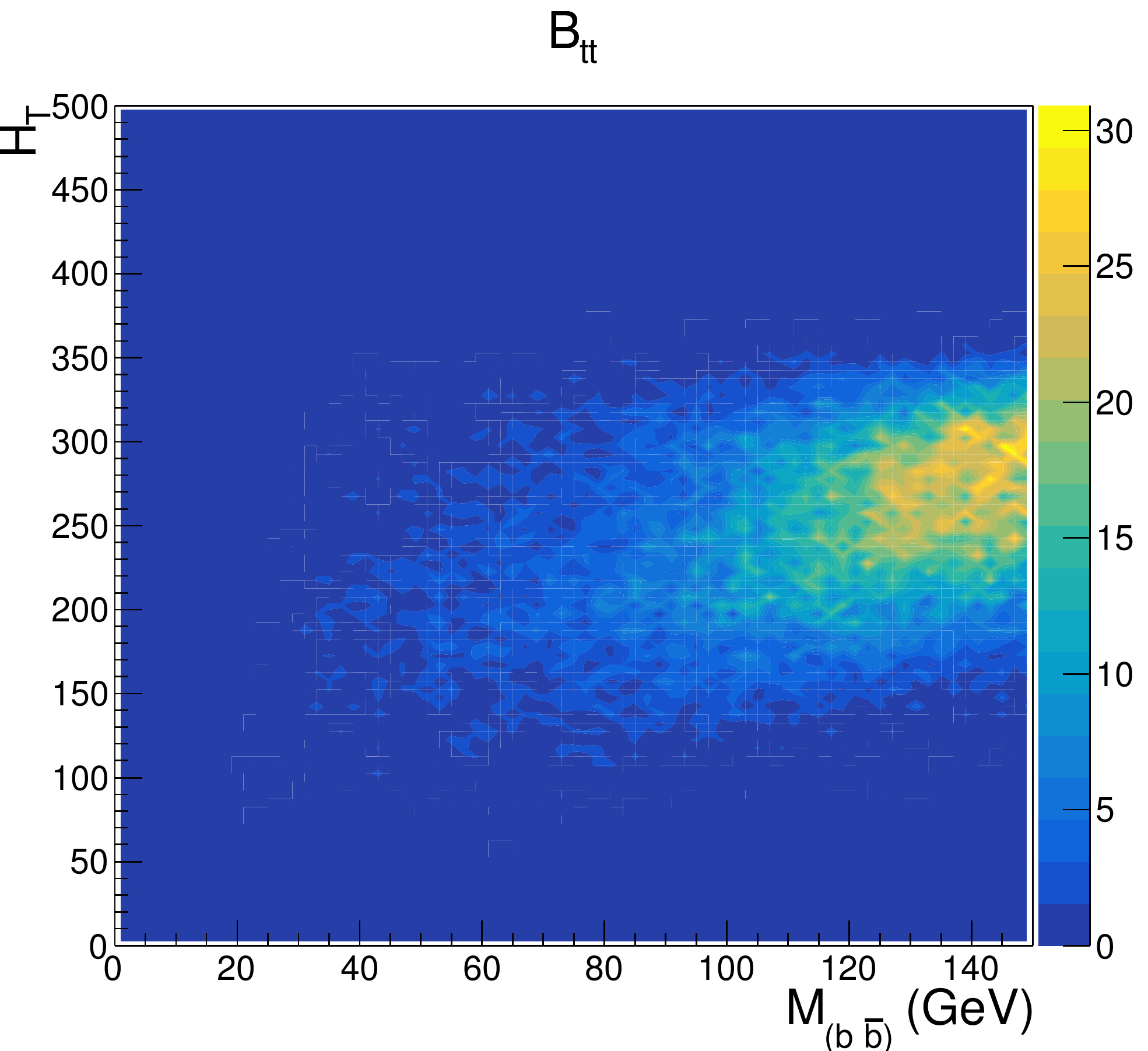}
\includegraphics[scale=0.16]{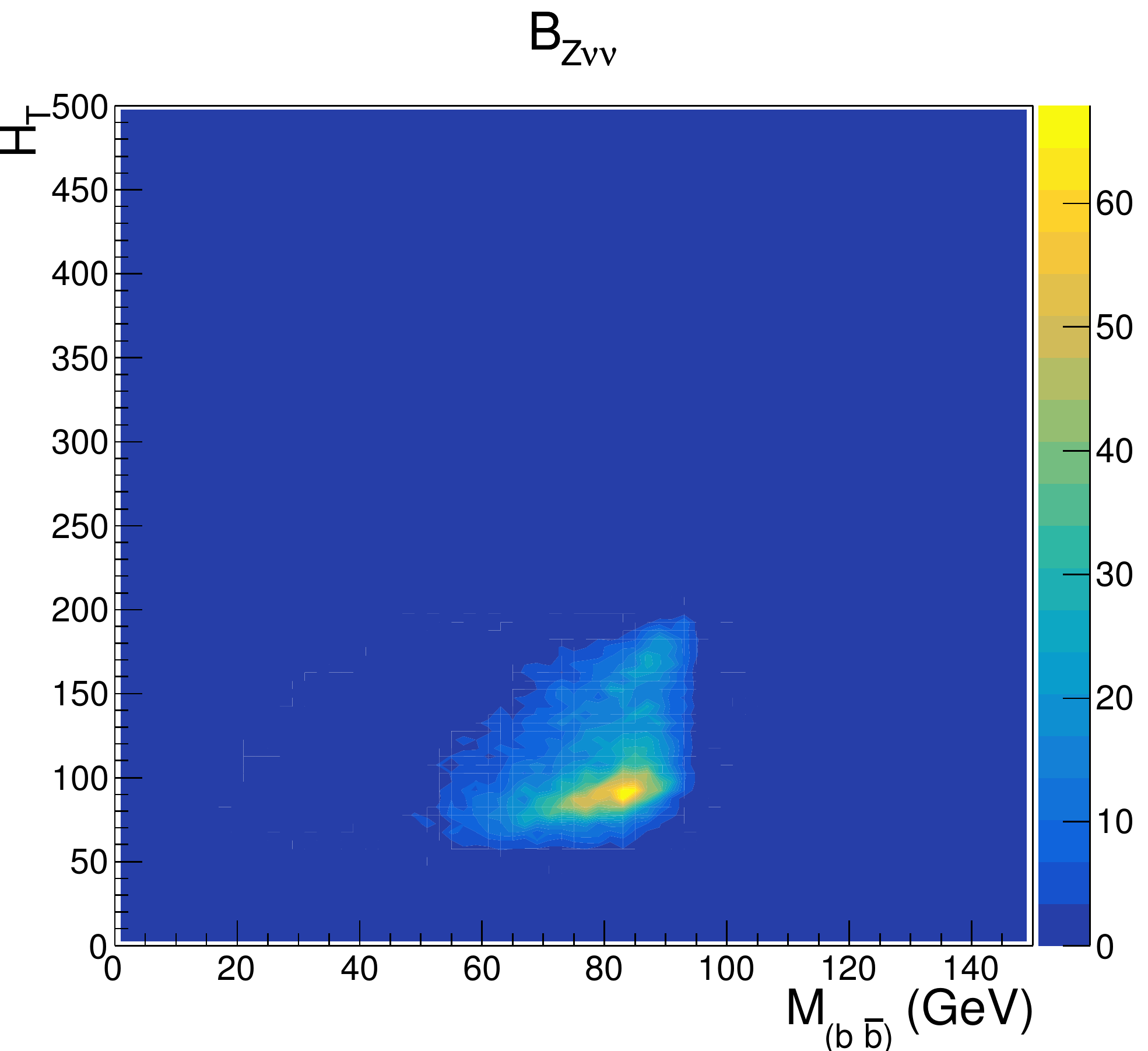}
\caption{  Normalized distributions of missing transverse energy (first row) and scalar transverse energy sum (second row) for signal versus reconstructed Higgs boson-mass from $b1$ and $b2$ ($M_{b,\bar b}$) with $\bar{c}_{HW} $=0.05 and relevant backgrounds processes. \label{fig3}}
\end{figure}

\begin{figure}
\includegraphics[scale=0.4]{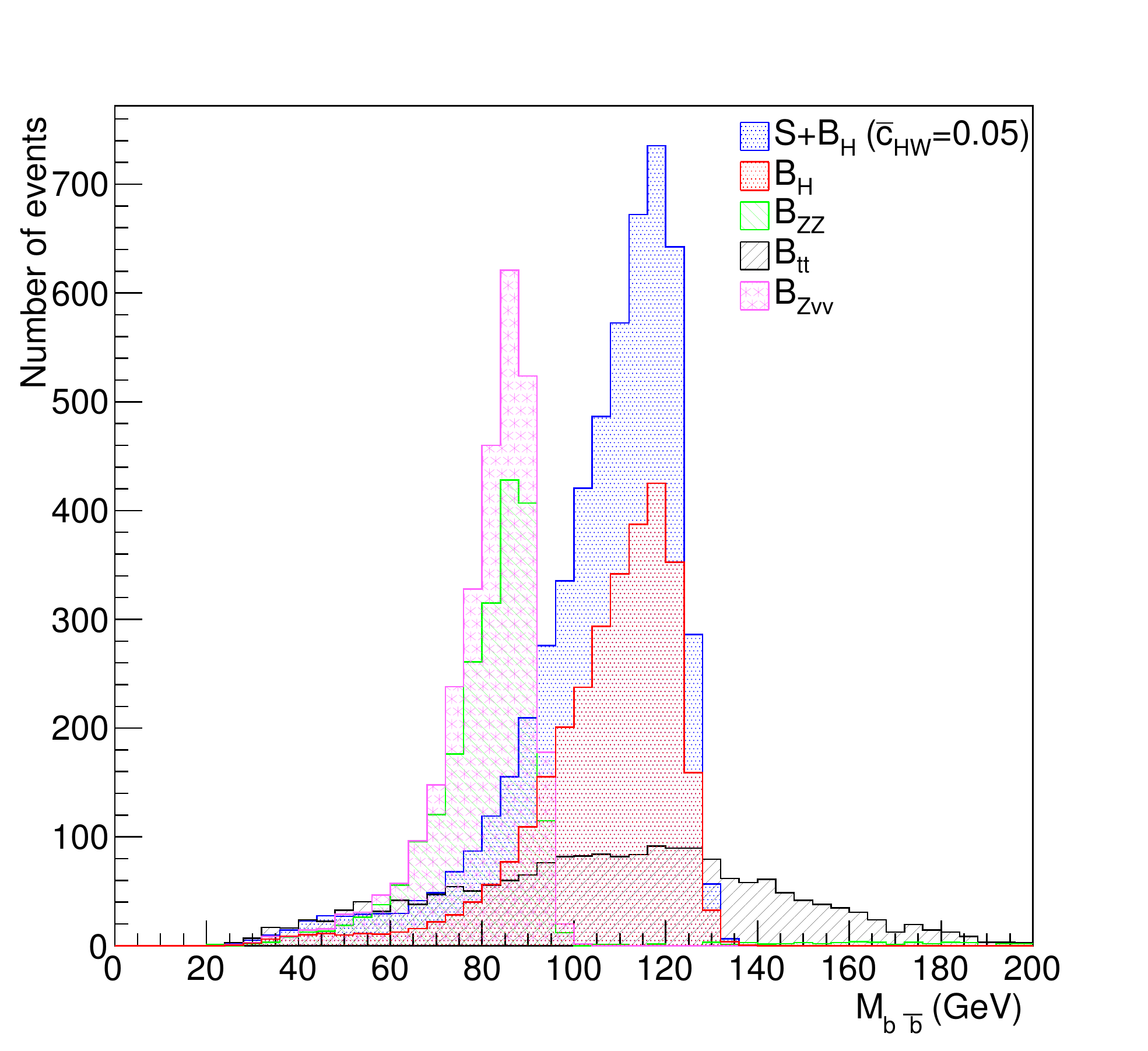}
\includegraphics[scale=0.4]{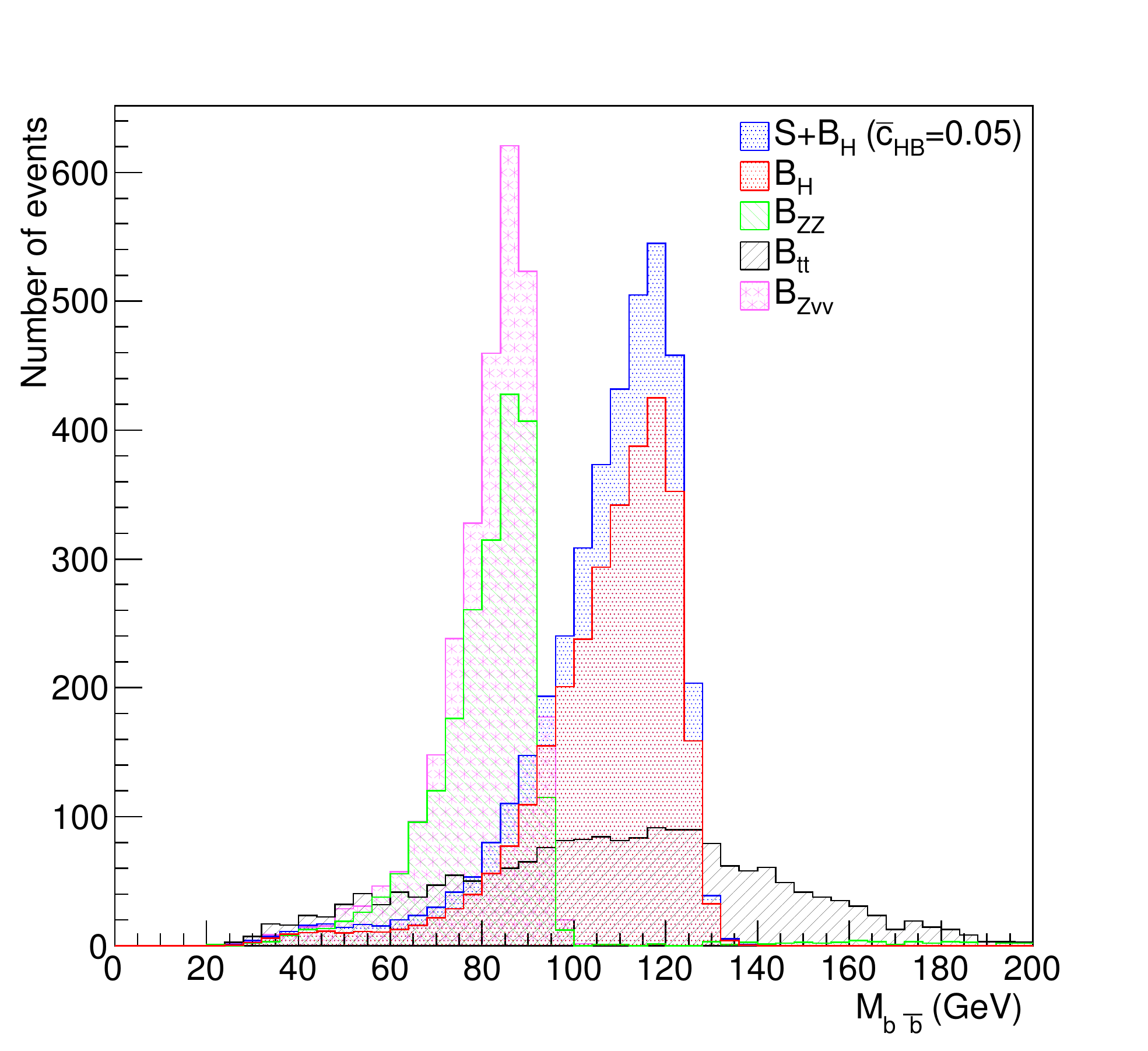}\\
\includegraphics[scale=0.4]{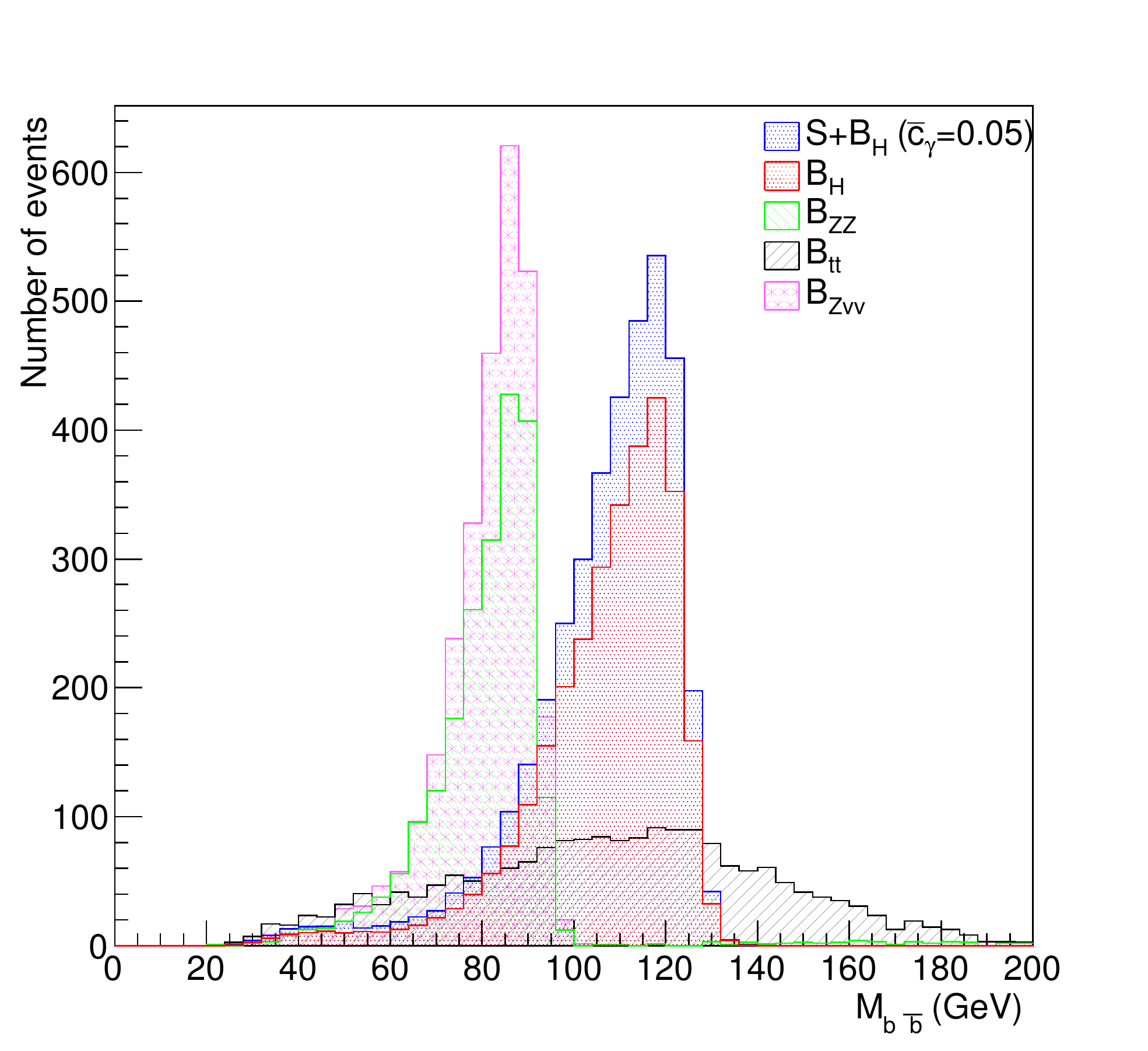}
\includegraphics[scale=0.4]{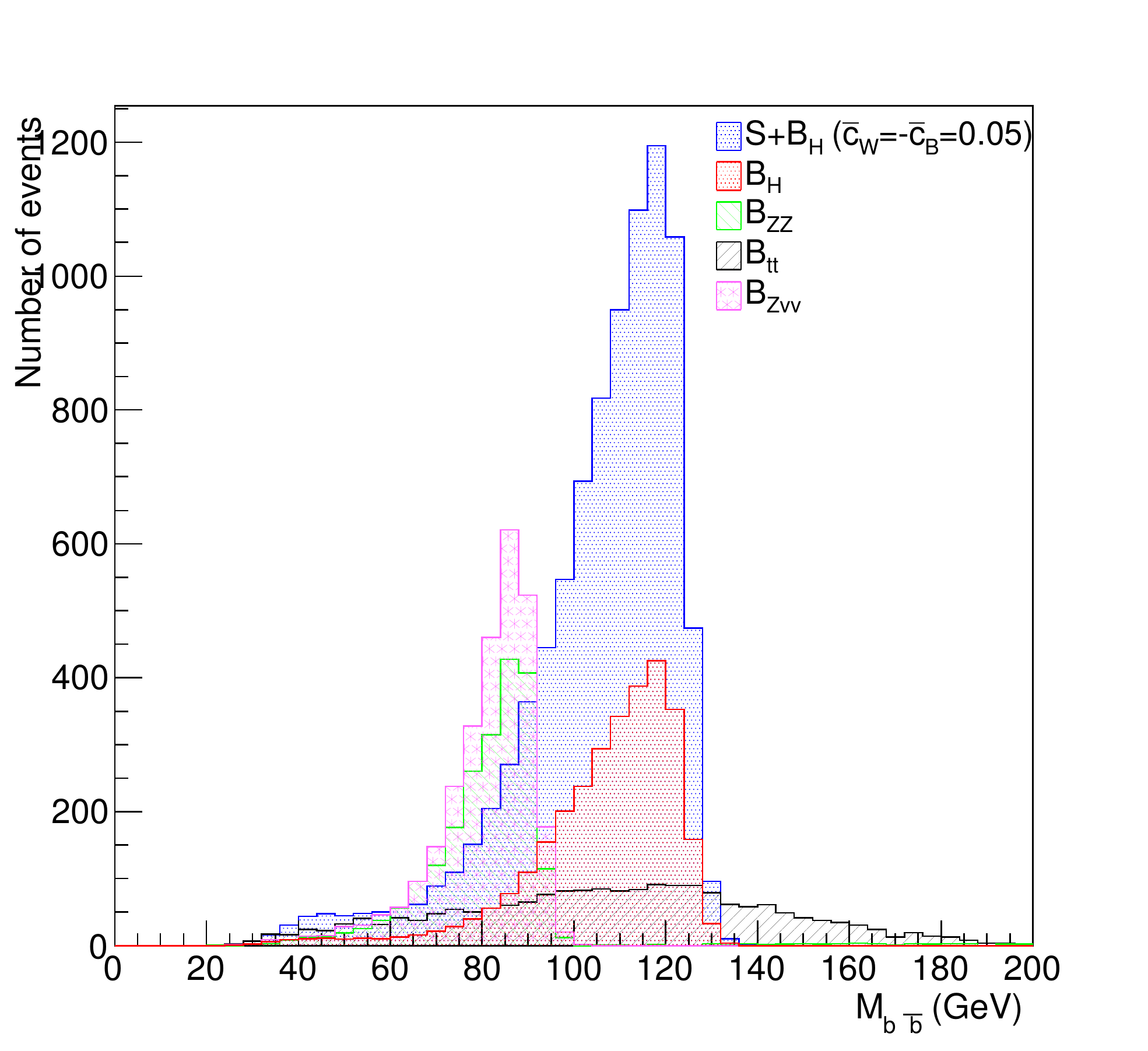}
\caption{ Normalized distributions of reconstructed invariant mass of Higgs-boson from $b\bar b$ for signal with $\bar{c}_{HW} $=0.05, $\bar{c}_{HB} $=0.05, $\bar{c}_{\gamma} $=0.05, $\bar{c}_{W}=- \bar{c}_{B}$=0.05   and relevant backgrounds processes. \label{fig4}}
\end{figure}

Requiring missing energy transverse (${\not}E_T$), no charged leptons and at least 2 jets with their transverse momenta ($p_T^j$) greater than 20 GeV and pseudo-rapidity ($\eta^j$) between -2.5 and 2.5 are the pre-selection of the event to be further analysed. The energy resolution of jets for $|\eta^j|\leqslant 3$ is assumed to be
\begin{eqnarray}
\frac{\Delta E_{jets}}{E_{jets}}=1.5\%+\frac{50\%}{\sqrt{E_{jets}(GeV)}}
\end{eqnarray}
The momentum resolution for jets as a function of $p_T^j$ and $\eta^j$  is 
\begin{eqnarray}
\frac{\Delta p_T^{j}}{p_T^{j}}=(1.0+0.01\times p_T^{j})\times 10^{-3}~~ \textrm{for}~~|\eta^j|\leqslant 1\\
\frac{\Delta p_T^{j}}{p_T^{j}}=(1.0+0.01\times p_T^{j})\times 10^{-2} ~~\textrm{for} ~~~1<|\eta^j|\leqslant 2.4
\end{eqnarray}
Jets are clustered with the anti-$k_t$ algorithm \cite{Cacciari:2008gp} using FastJet \cite{Cacciari:2011ma} where a cone radius is used as $R = 0.5$. In order to select the signal and background events, the following kinematic cuts and requirements are applied; 
\textbf{i)} requiring at least two jets tagged as the b-jet which significantly suppress the light-quark jet backgrounds. These two b-jets are used to reconstruct Higgs boson-mass. \textbf{ii)} One of the b-tagged jets with the highest $p_T$ is defined as $b1$ while the other is $b2$ with lower $p_T$. Fig.~\ref{fig2} shows $p_T$  distributions of $b1$ and $b2$ of signal (for $\bar{c}_{HW}$=0.05) and all relevant background processes versus reconstructed Higgs boson-mass from $b1$ and $b2$ ($M_{b,\bar b}$). As it can be seen in Fig.~\ref{fig2}, the $b_1$ with $p_T^{b1}>50$ GeV, $b_2$ with $p_T^{b2}>$30 GeV and pseudo-rapidity of the b-tagged jets to be $|\eta^{b1,b2}|\leqslant 2.0$ are considered to reduce $B_{ZZ}$ and $B_{Z\nu\nu}$. In ILD detector card, both b-tagging efficiency and misidentification rates are given as function of jet transverse momentum. For the transverse momentum of leading jet ($b1$) ranging from 50 GeV to 180 GeV, b-tagging efficiency is between 64\% and 72 \%, c-jet misidentification rate is 17\%-20\%, and misidentification rate of light jet 1.2\%-1.76\%. 
The missing transverse energy (${\not}E_T$) and scalar transverse energy sum ($H_T$)  for signal (for $\bar{c}_{HW}$=0.05) and all relevant background processes versus $M_{b,\bar b}$ are shown in Fig.~\ref{fig3}. \textbf{iii)} The missing transverse energy is required to be ${\not}E_T>30$ GeV to suppress the backgrounds  at low missing energy region. \textbf{iv)} Especially, to reduce $tt$ background process, the scalar transverse energy sum ($H_T$) is required to be 100 GeV $<H_T<$ 200 GeV. 
Normalized distributions of reconstructed invariant mass of Higgs-boson from $b\bar b$ for signal with $\bar{c}_{HW} $=0.05, $\bar{c}_{HB} $=0.05, $\bar{c}_{\gamma} $=0.05, $\bar{c}_{W}=- \bar{c}_{B}$=0.05  and relevant backgrounds processes are given in Fig.\ref{fig4}.\textbf{ v)} Finally, the reconstructed invariant mass of Higgs-boson from two b-jet is selected to be in the range 92 GeV $< M_{inv}^{rec}(b1,b2)< 136$ GeV. The kinematic distributions for each processes are normalized to the number of expected events which is defined to be the cross section of each processes times integrated luminosity with $L_{int}$=500 fb$^{-1}$. 

Effects of the cuts used in the analysis can be seen from the Table~\ref{tab2} which shows number of events after each cut. Requiring two b-tagged jets reduces the  $B_{ZZ}$, $B_{tt}$ and $B_{Z\nu\nu}$ backgrounds more than signal $S+B_H$ and background with same final state, $B_H$. Cut-2 effects on both signal and all relevant backgrounds, especially $B_{ZZ}$ and $B_{Z\nu\nu}$. ${\not}E_T$ cut decreases both $B_{tt}$ and $B_{ZZ}$ backgrounds while $H_T$ cut significantly suppresses $B_{tt}$ background. Final effect of the all cuts is approximately 15\%  for signal $S+B_H$ and $B_H$ background while 0.3\%-0.8\% for other relevant backgrounds. 
 \begin{table}
\caption{Number of signal and background events after applied kinematic cuts used for the analysis for $\bar{c}_{HW}$=0.05 with $L_{int}=500$ fb$^{-1}$  \label{tab2}}
\begin{ruledtabular}
\begin{tabular}{clcccccc}
Cuts & Definitions&$S+B_H$&$B_H$&$B_{ZZ}$&$B_{tt}$&$B_{Z\nu\nu}$ \\ \hline 
Cut-0 &  $N_j \geqslant 2$, lepton vetos, MET>0 with pre-selection cuts & 30432.5&19383.2&207847&211384&94405.6 \\
Cut-1 & two jets with $b$-tagging&8003.41&5035.96&10047.4&25636.2&6995.5  \\
Cut-2 & $p_{T}^{b1} > 50$ GeV, $p_{T}^{b2} > 30$ GeV and  $| \eta^{b1,\,b2}|\leqslant 2.0$&5862.8&3421.5&6662.6&24766.9&3205.9\\
Cut-3 & ${\not}E_T>30$ GeV&5548.4&3109.1&2705.4&8686.6 &3122.5 \\
Cut-4 & 100 GeV $<H_T<$ 200 GeV  &5407.9&3017.9&2158.1&1927.1&2823.3  \\
Cut-5 &100 GeV $< M_{inv}^{rec}(b1,b2)< 136$ GeV& 4211.6&2433.7&17.9&824.8&20.9 \\
\end{tabular}
\end{ruledtabular}
\end{table}

\section{sensitivity of  Higgs-gauge boson couplings}
\begin{figure}
\includegraphics[scale=0.8]{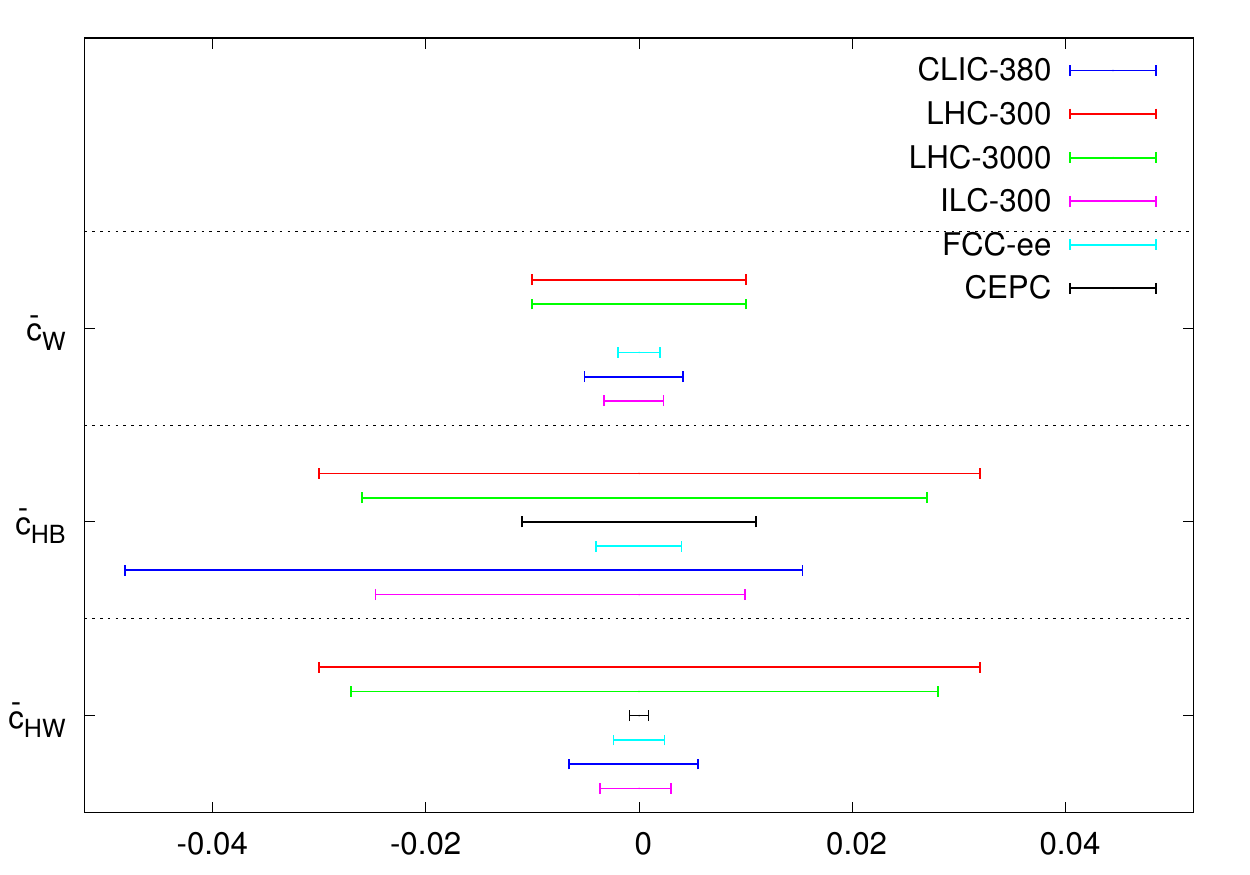} 
\caption{ Obtained allowed range (CLIC-380), LHC at 14 TeV center of mass energies for the integrated luminosity of 300 fb$^{-1}$ (LHC-300) and 3000 fb$^{-1}$ (LHC-3000) \cite{Englert:2015hrx}, ILC-300 at $\sqrt s=$ 500 GeV with $L_{int}=$300 fb$^{-1}$  \cite{Khanpour:2017cfq,Ellis:2015sca}, FCC-ee  for $L_{int}$=10 $ab^{-1}$ at $\sqrt s=$ 240 GeV \cite{Ellis:2015sca}, CEPC for $L_{int}$=5 $ab^{-1}$ at $\sqrt s=$ 240 GeV \cite{Ge:2016tmm} at 95\% C.L. for $\bar{c}_{HW} $, $\bar{c}_{HB} $, $\bar{c}_{W}=- \bar{c}_{B}$ coefficients. The limits are each derived with all other coefficients set to zero.   \label{fig5}}
\end{figure}

We calculate the sensitivity of the dimension-6 Higgs-gauge boson couplings in $e^+e^-\to \nu \nu H$ process  by
applying $\chi^{2}$ criterion with and without a
systematic error. The $\chi^{2}$ function is defined as follows
\begin{eqnarray}
\chi^{2} (\bar{c_i})=\sum_i^{n_{bins}}\left(\frac{N_{i}^{NP}(\bar{c_i})-N_{i}^{B}}{N_{i}^{B}\Delta_i}\right)^{2}
\end{eqnarray}
where $N_i^{NP}$ is the total number of events in the existence of effective couplings ($S$) and total backgrounds ($B_H$, $B_{ZZ}$, $B_{tt}$ and $B_{Z\nu\nu}$), $N_i^B$ is number of events of total backgrounds in $i$th bin of the invariant mass distributions of reconstructed Higgs boson, $\Delta_i=\sqrt{\delta_{sys}^2+\frac{1}{N_i^B}}$ is the combined systematic ($\delta_{sys}$) and statistical errors in each bin. So, the numerator in Eq.(7) equals to the number of extra events due to the presence of new operators.
In this analysis, we focused on $\bar{c}_{HB} $, $\bar{c}_{W}=- \bar{c}_{B}$ and $\bar{c}_{HW}$ couplings which are the
main coefficients contributing to $e^+e^-\to\nu \bar{\nu} H$ signal process. The 95\% Confidence Level (C.L.) limits including only statistical error on dimension-6 Higgs-gauge boson couplings at $\sqrt s$=380 GeV and $L_{int}$=500 fb$^{-1}$ (CLIC-380) are compared with the LHC at 14 TeV center of mass energies for the integrated luminosity of 300 fb$^{-1}$ (LHC-300) and 3000 fb$^{-1}$ (LHC-3000) \cite{Englert:2015hrx} in Fig.~\ref{fig5}. We see that CLIC-380 results would be significantly more sensitive to $\bar{c}_{HW}$ and somewhat sensitive to $\bar{c}_{W}=- \bar{c}_{B}$ whereas sensitivity to $\bar{c}_{HB} $ is comparable with expected LHC results. The prediction on the limits for the future lepton colliders; ILC \cite{Khanpour:2017cfq,Ellis:2015sca} of an integrated luminosity $L_{int}=$300 fb$^{-1}$ at the center of mass energy $\sqrt s=$ 500 GeV, FCC-ee \cite{Ellis:2015sca} for $L_{int}$=10 $ab^{-1}$ at $\sqrt s=$ 240 GeV, CEPC \cite{Ge:2016tmm} for $L_{int}$=5 $ab^{-1}$ at $\sqrt s=$ 240 GeV  are also shown in Fig.~\ref{fig5} .  In order to include the systematical uncertainties  we recompute the bounds. For example including a 10\% conservative systematic uncertainty, the constraint on $\bar{c}_{HW}$ is $[-0.07959; 0.02423]$. This bound is four times lower than the obtained limits without systematic uncertainties.


\section{Conclusions}
We have investigated the CP-conserving dimension-6 operators of Higgs boson with other SM gauge boson via $e^+e^-\to\nu \bar{\nu} H$ process using an effective Lagrangian approach at first energy stage of CLIC ($\sqrt s=380$ GeV, $L_{int}$=500 fb$^{-1}$). We have used leading-order strongly interacting light Higgs basis assuming vanishing tree-level electroweak oblique parameterize and flavor universality of the new physics sector. We analyzed only hadronic ($b\bar b$ ) decay channel of the Higgs boson including dominant background processes by considering realistic detector effect in the analysis. We have shown the kinematic distributions of b-jets in final state, missing transverse energy, scalar transverse energy sum and invariant mass distributions. Due to the fact that the signal final state consists of two neutrinos and two b-jets, the distributions of missing transverse energy and scalar transverse energy sum are performed for determining a cut-based analysis. We have obtained 95 \% C.L. limits on dimension-six operators analysing invariant mass distributions of two b-jets from Higgs decay in $e^+e^-\to\nu \bar{\nu} H$ signal process and the other dominant backgrounds. The $e^+e^-\to\nu \nu H$ process is more sensitive to $\bar{c}_{HW}$ couplings than the other dimension-six couplings at first energy stage of CLIC. Our results show that a CLIC with $\sqrt s=380$ GeV, $L_{int}$=500 fb$^{-1}$ will be able to probe the dimension-six couplings of Higgs-gauge boson interactions in $e^+e^-\to\nu \bar{\nu} H$  process especially for $\bar{c}_{HW}$ couplings at scales beyond the HL-LHC bounds while they become competitive with the $\bar{c}_{HB} $, $\bar{c}_{W}=- \bar{c}_{B}$ couplings. 

\begin{acknowledgments}
This work was partially supported by the Abant Izzet Baysal University Scientific Research Projects under the Project no: 2017.03.02.1137. H. Denizli's work was partially supported by Turkish Atomic Energy Authority (TAEK) under the grant No. 2013TAEKCERN-A5.H2.P1.01-24. The authors would like to thank to CLICdp group for the discussions, especially to Philipp G. Roloff for valuable suggestions in the CLICdp Working Group analysis meeting. Authors would also like to thank to L. Linssen for encouraging us to involve in CLICdp collaboration. 
\end{acknowledgments}

\end{document}